\documentclass [aps,twocolumn,showpacs,reprint]{revtex4-1}
\usepackage{graphicx}
\usepackage{textcomp}
\usepackage{psfrag}                    
\usepackage{float}
\begin{document}
\title{Modeling the complexity of  acoustic emission   during  intermittent plastic deformation: Power laws and multifractal spectra} 

\author{Jagadish Kumar $^1$ and  G. Ananthakrishna$^2$}

\affiliation{$^1$ Department of Physics, Utkal University, Bhubaneswar 751004, India \\ $^2$ Materials Research Centre, Indian Institute of Science, Bangalore 560012, India}

\begin{abstract}
Scale invariant power law distributions  for  acoustic emission signals are ubiquitous to several plastically deforming materials. However, power law distributions for the acoustic emission  energies are reported in distinctly different plastically deforming situations such as  hcp and,  fcc single and polycrystalline samples exhibiting smooth stress-strain curves,  and in  dilute metallic alloys exhibiting discontinuous flow. This is surprising  since the underlying dislocation mechanisms in these two types of deformations  are very different.  So far, there has been no models that predict  the power law statistics for  the discontinuous flow.  Furthermore,  the statistics of the acoustic emission  signals in   jerky flow is even more complex requiring multifractal measures for a proper characterization. There has been no model that explains the complex   statistics either. Here, we  address the problem of statistical characterization of  the acoustic emission signals associated with the three types of the Portevin-Le Chatelier bands. Following  our recently proposed  general framework for calculating acoustic emission, we set-up a wave equation for the elastic degrees of freedom with plastic strain rate as a source term.  The  energy dissipated during acoustic emission is represented by  the Rayleigh-dissipation function.   Using the plastic strain rate obtained from the Ananthakrishna model for the Portevin-Le Chatelier effect,  we  compute the acoustic emission signals associated the three  Portevin-Le Chatelier bands and the  L\"{u}ders like band. The so calculated acoustic emission  signals are used for further statistical characterization. Our results show that the model predicts power law statistics for all the acoustic emission signals associated with the three types of Portevin-Le Chatelier bands with the exponent values increasing  with increasing strain rate. The calculated multifractal spectra corresponding to the acoustic emission  signals associated with the three band types  has a maximum spread for the type C decreasing with type B and A.  We further show that the acoustic  emission  signals associated with L\"{u}ders like band also exhibits power law distribution and multifractality. 
\end{abstract}
\maketitle
\section{Introduction} 
It is well known that the motion of dislocations is inherently intermittent  with waiting periods at junctions where they are arrested  followed by near free flights between them. This feature is not reflected during  homogeneous yield phenomenon where the stress-strain ($\sigma-\epsilon$) curves are smooth. However, the intermittent behavior re-appears as serrations on the stress-strain curves when the diameter of the rod is decreased below a fraction of a micron  confirming the inherently intermittent character of dislocation motion \cite{Dimiduk09,Zaiser08}. On the other hand,  jerky flow or the  Portevin-Le Chatelier (PLC) effect is observed at macroscopic scales  when specimens of dilute alloys are subjected to constant strain rate deformation \cite{Anan07,Yilmaz11}. The intermittent deformation manifests itself as a series of serrations on the stress-strain curves in a range of temperatures and strain rates. A standard explanation is that the discontinuous flow is due to collective pinning and unpinning of dislocations from solute cloud. Clearly, the underlying dislocation mechanisms must necessarily be different in these two cases. 
 
Acoustic emission  (AE) technique has long been used as a tool to probe the dynamics of dislocations on finer scales  in several kinds of deformation studies, in particular, in the two cases mentioned above.  For instance,  AE  studies on the smooth homogeneous yield phenomenon exhibit  intermittent AE signals with its overall envelope exhibiting  a  peak just beyond the elastic regime decaying for larger strains  \cite{Dunegan69,Han11}. Recent AE studies for this case  throw light on  the origin of the  pulse like character of AE  signals \cite{Weiss97,Weiss07,Miguel01,Lebymf-pl-13}. The smooth $\sigma-\epsilon$ curves are then interpreted as resulting from the averaging process of the dislocation activity in the sample. 

Acoustic emission studies carried out for over five decades  have  established specific  correlations between the nature of the AE signals and the nature of stress-strain curves for  different situations \cite{Dunegan69,James71,Caceres87,Han11,Zeides90,Chmelik02,Chmelik07,Zuev08,Shibkov11}. 
The  nature of the AE signals  in the case of discontinuous flow  where the stress-strain curves display stress serrations is qualitatively different from that for the continuous homogeneous yield. For example, studies on the PLC effect have established specific types of correlations  between the nature of the AE signals  and the three  different types of  bands and the associated serrations \cite{James71,Zeides90,Chmelik02,Chmelik07,Lebymf09,Lebymf-Acta-12a,Lebymf-pl-13}.  For the uncorrelated static type C  bands that displays large amplitude serrations, the AE signals consist of  well separated AE bursts with  a low level AE background  while for the partially propagating type B bands exhibiting relatively small amplitude serrations, the AE bursts begin to overlap.  The AE signals  corresponding to the  fully propagating type A bands (with very small amplitude serrations)  is continuous with  bursts of AE  appearing only occasionally \cite{Zeides90,Chmelik02,Chmelik07,Lebymf-Acta-12a}. These characteristic features have been captured in our recent work \cite{Jag11,Jag15}.  Similar correlations exist for the L{\"u}ders band \cite{Chmelik02,Chmelik07,Zuev08,Shibkov11}, another type of propagative instability \cite{Neuhausser83}. Furthermore, the AE signals for  the L{\"u}ders  band are  different from that for the three  PLC bands \cite{Anan07,Yilmaz11}, a feature that  is also captured by our model \cite{Jag15}. 

In the case of propagative instabilities, collective dislocation processes  govern the nature of the bands and the associated stress-strain curves.  While the AE associated with both the continuous yield and the discontinuous yield  consist of a sequence of intermittent bursts of acoustic energy,  the AE technique is  not only able to distinguish these two types of AE patterns  but also to quantify them.  Indeed, the recent progress in experimental techniques  with high degree of  resolution and accuracy have helped to establish   quantitative characterization of the AE signals corresponding to the different types of stress-strain curves \cite{Weiss97,Miguel01,Chmelik02,Chmelik07,Leby-pl-Acta-12,Shashkov-pl-12,Lebymf-pl-13}. 

One characteristic feature of  AE signals  is the scale free statistics of the amplitude (or energy) in  widely different physical systems such as  volcanic activity \cite{Diodati91},  microfracturing process \cite{Petri94}, thermal cycling of martensites \cite{Planes95,Rajeevprl,Kalaprl}, peeling of an adhesive tape \cite{Cicc04,Rumiprl,Jag08} and plastic deformation of materials exhibiting homogeneous yield \cite{Weiss97,Miguel01,Lebymf-pl-13} as well as discontinuous flows \cite{Leby-pl-Acta-12,Shashkov-pl-12,Lebymf-pl-13}. In the context of plastic deformation,   the statistics of the recorded intermittent AE signals obtained from an uniaxial compressive creep of ice crystals,  single and polycrystalline samples of hcp and fcc structures \cite{,Lebymf-pl-13} follow power law distributions.  Surprisingly,  even in the  case of the PLC effect, power law distributions for the  AE energies are reported for all the PLC  bands \cite{Lebymf-Acta-12a, Shashkov-pl-12,Lebymf-pl-13}. Interestingly, while the scaling region for the AE energies corresponding to the type A bands  has the least scaling regime, it is only for the type A band that  power law distributions for the magnitudes and durations of stress drops have been reported  \cite{Anan99,Leby95,Bhar-Acta-02,Leby-pl-Acta-12}.  

The fact that power law distributions of the AE signals are seen in these two cases (of homogeneous and discontinuous flows) with distinctly different dislocation dynamics  is surprising. The power law statistics for the homogeneous yield has been predicted in 2-D and 3-D simulations \cite{Miguel01,Csikor07}. The underlying physical mechanism attributed is the formation of  dislocation network that is driven to the edge of criticality. In this state, some proportion of the network is poised at and just below the criticality. Then, bursts of AE are attributed to the coherent unpinning of those dislocation segments at criticality.  Once unpinned, these segments get arrested falling below the critical state with segments below the criticality pushed to  the threshold, much like the sand-pile model scenario \cite{Bak88}. Clearly, this kind of explanation cannot hold for the power law statistics of the AE signals associated with the PLC  bands  since these band types are attributed to collective unpinning of dislocations.  {\it To the best of our knowledge, there are no models or simulations that predict  the power law statistics of the AE signals in the case of the PLC effect.}   Thus, our primary  objective is to examine if the calculated model AE signals  associated with the three PLC bands \cite{Jag15}  exhibit  power law statistics.

Even more interesting is the fact that the statistical properties of the AE signals from  the PLC bands are even more complex   requiring  a distribution of scaling exponents for a proper characterization unlike a single exponent required for characterizing a power law distribution  \cite{Lebymf-Acta-12a,Shashkov-pl-12,Lebymf-pl-13}. Indeed, the statistical properties  of such complex sets are mathematically characterized by  a continuum of generalized Renyi dimensions $D_q$ parametrized by a parameter $q$  \cite{Renyi,Halsey86,Chhabra89}. Alternately, they can be described as an  interwoven sets of Hausdorff (fractal) dimensions  $f(\alpha)$ having a singularity strength $\alpha$ \cite{Halsey86,Chhabra89}. So far, there is no model that  reports the generalized dimensions $D_q$ or the singularity spectrum $f(\alpha)$ for the AE energies associated with the three PLC bands. Thus,  our second  objective is to examine if the calculated  AE signals  for the PLC bands \cite{Jag11,Jag15} exhibits multifractality.

Towards this end, we follow the recently  proposed  framework for calculating AE for any plastic deforming situation. We use the Rayleigh dissipation function to represent the energy dissipated during acoustic emission. The method involves setting-up a discrete set of wave equations  for the elastic strain with plastic strain rate as a source term. The plastic strain source term is itself calculated using the Ananthakrishna (AK)  model for the PLC effect since the model predicts all the generic features of the PLC instability including the three band types \cite{Jag11,Jag15}. This model  also predicts L\"{u}ders like bands.  We demonstrate  that the model AE energy bursts corresponding to the three types PLC bands and  L\"{u}ders like band  predict both  power law distributions and multifractal spectra.

\section{Theoretical framework for acoustic emission during plastic deformation}

We briefly recall the theoretical framework developed  for calculating AE  in our recent papers \cite{Jag11,Jag15}. The basic idea is that acoustic emission (transient elastic waves) are generated due to the abrupt motion of dislocations. In mathematical terms, this translates to setting-up a wave equation for the elastic degrees of freedom with  plastic strain rate as a source term. At this point, {\it the equation is general enough to be applicable to any plastic deformation} since the plastic strain rate source term has no specific reference to the nature of plastic deformation. Therefore,  to be applicable to a specific  a plastic deformation, we need to devise a suitable model that captures the characteristic features of the  plastic deformation. In the present context, the minimum requirement is that the model should predict  all the generic features of the PLC effect including  the three types of bands and the associated serrations. The AK model  meets this requirement since the model  captures most generic features of the PLC effect including the band types \cite{Anan07,Anan82,Bhar02,Bhar03,Bhar03a,Anan04,Ritupan15}. The model also predicts  L\"{u}ders like band.

\subsection{Approach}
As stated above, the abrupt motion of dislocations sets-off elastic disturbance,  which then activates the dissipative forces that tend to oppose their growth  so that mechanical equilibrium is restored. Then, the dissipative energy   \cite{Rumiepl,Rajeevprl,Kalaprl,Rumiprl,Jag08} during AE represented by the  Rayleigh dissipation function ${\cal R}_{AE}$ \cite{Land} either in terms of the elastic strain $\epsilon_e$ or in terms of displacement field $u(y,t)$. Then, we have  
\begin{equation}
{\cal R}_{AE} = {\eta\over2}\int\Big[{\partial \epsilon_e\over \partial t} \Big]^2 dy ={\eta\over2}\int \Big[{\partial \dot u(y,t) \over \partial y} \Big]^2dy.
\label{P-dissip}
\end{equation}
where we have used the elastic strain  defined by $\epsilon_e(y,t) = \frac{\partial u(y,t)}{\partial y}$.  Here, $\eta$ the damping co-efficient.  Then, since  ${\cal R}_{AE} \propto \dot \epsilon_e^2(y)$, we interpret  $ {\cal R}_{AE}$ as the acoustic energy dissipated during the abrupt motion of  dislocations \cite{Rumiepl,Rajeevprl,Kalaprl,Rumiprl,Jag08,Ritupan14}.

The total energy of a   one dimensional crystal is the sum of  the kinetic energy, the potential energy and the gradient potential energy \cite{Rajeevprl,Kalaprl,Ritupan14}. The kinetic energy is given by $T =  {\rho\over 2} \int  [\frac{\partial u(y,t)}{\partial t}]^2 dy$ where $\rho$ is the linear mass density. In our work, we have chosen to use mass per unit volume  instead of linear mass density  so that  the elastic constant $\mu$ appears naturally  in the expression for the  potential  energy,  given by  $ V_{loc}={\mu\over2} \int\Big[{\partial u(y,t) \over \partial y} \Big]^2 dy$. (Note that the machine equation describing constant strain rate condition contains the elastic constant.) The gradient  potential energy is given by $V_{grad}=\frac{D}{4}\int\Big[{\partial^2 u(y,t) \over \partial y^2} \Big]^2 dy.$ Here $D$ is the gradient energy coefficient. Then, using the Lagrangian ${\cal L}= T-V_{loc}-V_{grad}$ in the Lagrange equations of motion, we get  
\begin{equation}
\rho \frac{\partial^2 u(y,t)}{\partial t^2} = \mu \frac{\partial^2 u(y,t)}{\partial y^2} +\eta \frac{\partial^2 \dot u(y,t)}{\partial y^2}- D \frac{\partial^4 u(y,t)}{\partial y^4}.
\label{Dwave}
\end{equation}
Noting that strain $\epsilon(y,t) $ is the  natural variable used in plastic deformation, the  wave equation in the strain variable is obtained by differentiating Eq. (\ref{Dwave}) with respect to $y$. Then, we have 
\begin{eqnarray}
\rho \frac{\partial^2 \epsilon_e}{\partial t^2} = \mu \frac{\partial^2 \epsilon_e}{\partial y^2} +\eta \frac{\partial^2 {\dot \epsilon_e}}{\partial y^2}- D \frac{\partial^4 \epsilon_e}{\partial y^4}.
\label{wave}
\end{eqnarray}
Equation (\ref{wave}) describes sound waves in the absence of dislocations. However, during plastic flow, since transient elastic waves are triggered by the abrupt motion of dislocations, we include  plastic strain rate $\dot\epsilon_p(y,t)$ as a source term in Eq. (\ref{wave}). Then, the relevant inhomogeneous wave equation describing the AE  process takes the form 
\begin{eqnarray}
\rho \frac{\partial^2 \epsilon_e}{\partial t^2} = \mu \frac{\partial^2 \epsilon_e}{\partial y^2} - \rho  \frac{\partial^2 \epsilon_p}{\partial t^2}+\eta \frac{\partial^2 {\dot \epsilon_e}}{\partial y^2}- D \frac{\partial^4 \epsilon_e}{\partial y^4}.
\label{comp_wave}
\end{eqnarray}
Here $c= \sqrt{\mu/\rho}$ is the velocity of sound. Note that $\dot\epsilon_p(y,t)$ is a function of both space and time and hence contains full information of any possible  heterogeneous character of the deformation. Clearly, the plastic strain rate source term needs to be {\it calculated independently} by  setting-up appropriate evolution equations for suitable types of dislocation densities for the desired plastic deformation. For the problem under consideration, we shall use the AK model for the PLC effect.

Solution of Eq. (\ref{comp_wave}) requires that we specify the initial and boundary conditions appropriate to the problem. Consider the commonly employed constant strain rate mode of deformation. This condition is enforced  by fixing one end of the sample  and applying a  traction at the other end. However,  it is important to ensure that the boundary conditions  on  Eq. (\ref{comp_wave}) be consistent with those imposed on the evolution  equations for the dislocation densities  (subject to  constant strain rate condition).  However, the values of  the dislocation densities at the boundaries are determined by physical considerations and therefore the  plastic strain rate  $\dot\epsilon_p(y,t)$ at the boundary  points  need not necessarily be  consistent with those  imposed on  Eq. (\ref{comp_wave}). (Note that the sum of  the elastic and plastic  strain rates should be equal to the imposed strain rate.) More importantly, in real samples,  the machine stiffness gripping the ends of the sample is higher than that for the sample. This condition is  not easy to include in the continuous form of the wave equation, i.e.,  Eq. (\ref{comp_wave}).  On the other hand, this condition can be easily implemented in the equivalent discrete set of wave equations given in the Appendix (see also Ref. \cite{Jag15}).  Such a discrete set of  wave  equations for $\epsilon_e(j,t)$, j =1 to N allows us to  make a distinction between points  within the sample and those at the boundary. Note that numerical solution of the evolution equations for the dislocation densities is obtained by solving them  on a grid of $N$ points corresponding to a sample length $L$ by fixing one end  and pulling the other end at a constant strain rate.  Then, the  plastic strain rate computed for each spatial point can be directly used as source terms in the discrete set of $N$ wave equations. The method also brings clarity to the boundary conditions.  A brief outline of the discrete set of wave equations   is given in the Appendix (Eqs. (\ref{waveqn_discret1}-\ref{waveqn_discret99}) )). For details see Ref. \cite{Jag15}.

\subsection{The Ananthakrishna model for the Portevin-Le Chatelier effect}
Equation (\ref{comp_wave}) (or Eqs. (\ref{waveqn_discret1}-\ref{waveqn_discret99}) of the Appendix) require that we supply the source term ${\dot \epsilon}_p(y,t)$ to compute the AE signals  associated with the PLC bands. This then can be used for  further statistical analysis. For the sake of completeness, we  briefly recapitulate the AK model  \cite{Anan07,Bhar02,Bhar03,Anan04,Anan82,Ritupan15} for the PLC effect  that was also used in Refs. \cite{Jag11,Jag15} for  calculating AE corresponding to  the three PLC bands.  

We begin with a brief summary of  the salient features of the PLC effect. The PLC effect is characterized by three types of  bands  \cite{Anan07,Yilmaz11} and the  associated stress serrations observed with increasing strain rate.  At low applied strain rates ${\dot \epsilon}_a$,  randomly nucleated static type C bands are seen with large amplitude nearly regular serrations.  With increase in ${\dot \epsilon}_a$, the hopping type B bands are seen   that exhibit  more irregular smaller serrations.  Finally, at high ${\dot \epsilon}_a$, the continuously propagating type A bands are observed. The associated serration amplitudes are considerably smaller.   These different types of PLC bands has been shown to  represent distinct correlated states of dislocations in the bands \cite{Anan07,Bhar02,Bhar03,Bhar03a,Anan04,Ritupan15}. 

We will use the AK model for the PLC effect since it predicts all the characteristic features of the phenomenon such as the existence of the instability within a window of strain rates, the negative strain rate behavior \cite{Anan82,Rajesh} and the three types of band types  C, B and A \cite{Anan07,Bhar02,Bhar03,Anan04,Anan82,Ritupan15}.  The model also predicts several dynamical features reported in the analysis of the experimental time series such as the existence of chaotic stress drops \cite{Anan83}, which has been subsequently verified \cite{Noro97,Anan99,Bhar01}. Inclusion of spatial degrees of freedom automatically predicts  the  cross-over dynamics from low dimensional chaos to infinite dimensional power law state of stress drops reported in the analysis of  experimental stress-strain series \cite{Noro97,Anan99,Bhar01,Bhar02,Bhar03,Bhar03a,Anan04}. In addition,  the AK model also predicts L\"uders like bands \cite{Jag15}.

The basic idea of the model is that all the qualitative features of the PLC effect emerge from a nonlinear interaction of  a few collective degrees of freedom, namely, the densities of the mobile, immobile and the dislocations with cloud of solute atoms denoted respectively by $\rho_m(x,\tau)$, $\rho_{im}(x,\tau)$ and    $\rho_c(x,\tau)$.  The evolution equations are
\begin{eqnarray}
\nonumber
\label{X-eqn}
\frac{\partial \rho_m}{\partial t'} &=& -\beta \rho_m^2 - f\beta \rho_m\rho_{im} -\alpha_m \rho_m + \gamma\rho_{im}  \\ 
&+ &\theta v_0 \big[\frac{\sigma_{eff}}{\sigma_y}\big]^m \rho_m + \frac{\Gamma\theta v_0}{\rho_{im}} \frac{\partial^2 }{\partial x^2}\big[\frac{\sigma_{eff}}{\sigma_y}\big]^m  \rho_m, \\
\label{Y-eqn}
\frac{\partial \rho_{im}}{\partial t'} &=& \beta \rho_m^2 - \beta \rho_m\rho_{im} -\gamma\rho_{im} + \alpha_c\rho_c,\\
\label{Z-eqn}
\frac{\partial \rho_c}{\partial t'} &= & \alpha_m\rho_m - \alpha_c\rho_c.
\end{eqnarray}
Equations (\ref{X-eqn}-\ref{Z-eqn}) have been discussed in a number of earlier publications and the details can be found in Refs. \cite{Anan07,Anan82,Bhar01,Bhar02,Bhar03,Bhar03a,Anan04,Ritupan15,Jag15}. For the sake of completeness, here, we provide a brief description of each of the dislocation mechanisms.

The first  term in Eq. (\ref{X-eqn}) refers to the formation of dipoles. This term acts as source term for  $\rho_{im}$. The second term refers to the annihilation of a mobile dislocation with an immobile one. $\beta$ has a dimension of the rate of the area swept by a dislocation. This is a common loss term to both $\rho_m$ and $\rho_{im}$.    The third term $\alpha_m\rho_m$ in Eq. (\ref{X-eqn}) corresponds to solute atoms diffusing to mobile dislocations temporarily arrested at immobile (or forest) dislocations. $\alpha_m$  can be expressed in terms of the solute concentration  $c$ at the core of the dislocations, $D_c$ the diffusion constant of the solutes and $\lambda$ an effective attractive length scale  for the solute diffusion. Then, $\alpha_m= D_c (T) c/\lambda^2$ (see \cite{Ritupan15}).  The fourth term $\gamma \rho_{im}$ is the reactivation  of the fraction of $\rho_{im}$ that has been immobilized due to solute pinning (see below). 

The loss term $\alpha_m \rho_m$ Eq. (\ref{X-eqn}) is a gain term for $\rho_c$. We consider those mobile dislocations  that start acquiring solute atoms as dislocations with solute atoms $\rho_c$.   As dislocations progressively acquire solute atoms at a rate $\alpha_c$, they are eventually immobilized, at which point they are considered as $\rho_{im}$. Then, the loss term $\alpha_c \rho_c$ in Eq. (\ref{Z-eqn}) is a source term  for  $\rho_{im}$ in Eq. (\ref{Y-eqn}). (Note that $1/\alpha_c$ represents the aging time.)  {\it  Thus, $\rho_{im}$ includes dislocations that are pinned by solute atmosphere as well.} Therefore, the loss term $\gamma \rho_{im}$ in Eq. (\ref{Y-eqn}) is considered to represent the unpinning of that fraction of  $\rho_{im}$ immobilized by  the  solute atoms.  

The fifth term in Eq. (\ref{X-eqn}) represents the rate of multiplication of dislocations due to cross-slip given by $ \theta v_m(\sigma_{eff}) \rho_m = \theta v_0[\sigma_{eff}/\sigma_y]^m \rho_m$, where $m$ is a velocity exponent. Here  $ v_m(\sigma_{eff})$ is the mean velocity of mobile dislocations taken to have a power law dependence on the effective stress $\sigma_{eff} = \sigma_a- h \rho_{im}^{1/2}$ \cite{Neuhausser83}. $\sigma_y$ is the yield stress and   $h\rho_{im}^{1/2}$ the back stress. ($h=\alpha G b$  where $\alpha \sim 0.3$, $G$ the shear modulus and $b$ the magnitude of the Burgers vector.) 

Several  types of spatial couplings such as solute diffusion to dislocations,  double cross-slip, long range elastic interaction between dislocations, compatibility stresses between grains, Bridgman factor (bending moments) and correlated motion of dislocation glide due to long range forces have been proposed (see pages 130-133 of Ref. \cite{Anan07}). In the context of theoretical modeling, even the long range spatial couplings such as the long range elastic interaction between dislocations,  Bridgman factor and correlated motion of dislocations have been reduced to diffusive type of coupling to the leading order \cite{Anan07}. In our model, the most natural spatial coupling (the sixth term in Eq. (\ref{X-eqn}))  arises from the double cross-slip process that allows dislocations to move into neighboring spatial elements.  Note that  $1/\rho_{im}$ factor prevents the occurrence of cross-slip into regions of high back stress. (See \cite{Anan04,Anan07} for details of the derivations for different types of  diffusive couplings.) 

These equations are coupled to the machine equation that enforces the constant strain rate  condition 
\begin{equation}
\frac{d\sigma_a}{dt'} = E^* \big[{\dot\epsilon}_a -\frac{b}{L}\int_0^L  v_0 \big[\frac{\sigma_{eff}}{\sigma_y}\big]^m \rho_m dx\big] = E^*[{\dot\epsilon}_a - \dot\epsilon_p(t')].
\label{S-eqn}
\end{equation}
Here, $E^*$ is the  effective modulus of the machine and the sample, and $L$ the length of the sample.   

Equations (\ref{X-eqn}-\ref{S-eqn}) are solved by using an adaptive step size differential equation solver (ode15s MATLAB solver).  The model parameters fall into two types, one experimental and the other theoretical. In our approach, we can adopt the experimental parameters. Theoretical parameters correspond to parameters associated with dislocation mechanisms used in the model. The instability domain for various parameters have been determined in a number of publications \cite{Anan07,Anan82,Bhar01,Bhar02,Bhar03,Bhar03a,Anan04,Ritupan15,Jag15}.   The parameters used for calculation of AE signals  (adopted from Ref. \cite{Jag15}) are given in Table \ref{T1}. 
\begin{table}[!h]
\caption{Parameter values used for the AK model for computation of the acoustic emission. }
\label{T1}
\begin{center}
\begin{tabular} {|c |c |c |c |c |}
\hline 
$E^* (GPa) $ & $ \sigma_y(GPa) $ & $\alpha_m(s^{-1}) $ & $\alpha_c(s^{-1}) $ & $ v_0 (ms^{-1}) $ \\
\hline
$48$  & $0.2$ & $0.8$ & $0.08$ & $ 10^{-7} $  \\ [0.1ex]
\hline
$\gamma (s^{-1})$ & $  f$ & $m $ & $ \beta (m^{2}s^{-1}) $ & $\Gamma$ \\
\hline
$5\times 10^{-4}$  & $1$ & $3$ & $5 \times 10^{-14}$ & $10^{12}$ \\ [0.5ex]
\hline
\end{tabular}
\end{center}
\end{table} 

\section{Power law statistics for the acoustic energy during jerky flow}

As stated in the introduction, the discrete set of wave equations for $\epsilon_e(j,t)$, j=1 to N given by  Eqs. (\ref{waveqn_discret1})-(\ref{waveqn_discret99}) provide the general framework for calculating the AE signals provided the plastic strain rate source term can be calculated. For the present case, ${\dot\epsilon}_p$ is calculated by using the AK model equations that reproduce all the generic features of the PLC effect. The underlying physical mechanism  subsumed in the AK model is the collective pinning and unpinning of dislocation from solute clouds.  Clearly, the unpinning mechanism  acts as  a  source generating the AE signals. 

The numerical steps adopted for computing  AE signals is detailed in Ref. \cite{Jag15}.   The first step is to  solve  Eqs. (\ref{X-eqn}-\ref{S-eqn})  on a  grid of $N$ points  for the entire  time interval and obtain ${\dot\epsilon}_p(k,t_i')$ using a fixed or variable time step $\delta t'$. The so computed ${\dot\epsilon}_p(k,t_i')$ can then be used as a source term in the $N$ discrete wave equations Eqs. (\ref{waveqn_discret1})-(\ref{waveqn_discret99}). However, noting that the discrete wave equations need to be integrated at a much finer time steps, say $\delta t$, we first obtain the corresponding interpolated values of ${\dot\epsilon}_p(k,t_i)$ at these time steps, which are then used for solving  Eqs. (\ref{waveqn_discret1})-(\ref{waveqn_discret99}).  This gives the elastic strain rate, which is used to  compute the AE energy dissipated using Eq. (\ref{P-dissip}). 
 
It should be mentioned here that above method of calculating  AE  is only approximate since we use the equilibrated value of stress (Eq. \ref{S-eqn} ) to obtain the plastic strain rate ${\dot\epsilon}_p(k,t')$. The method is akin to adiabatic schemes. However, it is possible to solve the AK model equations Eqs. (\ref{X-eqn})-(\ref{Z-eqn}) together with Eqs. (\ref{waveqn_discret1})-(\ref{waveqn_discret99}) to obtain the elastic strain and hence the instantaneous stress along with the plastic strain rate. The so computed stress will in general be different from the equilibrated value used in the approximate scheme. 

\subsection{Nature of the acoustic emission signals accompanying the PLC bands}
Recall that in our framework, acoustic emission is triggered by the plastic strain rate source term. The latter carries the physical information about the abrupt motion of dislocations.   It is therefore useful to briefly recall some results from our earlier work \cite{Ritupan15} on the nature of the stress serrations associated with the three band types and their correlations  with the nature of AE patterns \cite{Jag15}. At low strain rates, the uncorrelated static type C bands are seen with nearly regular large amplitude serrations.  As $\dot\epsilon_a$ is increased we see the  hopping type B bands.  The serrations are  more irregular and  are also of smaller  magnitude.  One important  feature predicted by the model relevant to the AE studies  is the correlation between band propagation property and small amplitude serrations (SASs), and its influence on the AE pattern.  It was recently demonstrated that {\it band propagation induces small amplitude serrations that are bounded on both sides by large amplitude stress drops \cite{Ritupan15}.} The  latter were  found to be  well correlated  with the nucleation and stopping of the band. As we increase $\dot\epsilon_a$, the extent of propagation increases with a concomitant increase of the SASs.  At high $\dot\epsilon_a$ (type A bands), occasional large amplitude stress drops accompany the  SASs, identified with the band reaching the boundaries. These features are consistent with experimental studies \cite{Ranc05,Jiang07}.  Another feature that is relevant for the AE studies is the fact that the mean stress level of these SASs increases or decreases, though this change  is small. 

\begin{figure}[!h]
\vbox{
\begin{center}
\includegraphics[height=4.0cm,width=7.9cm]{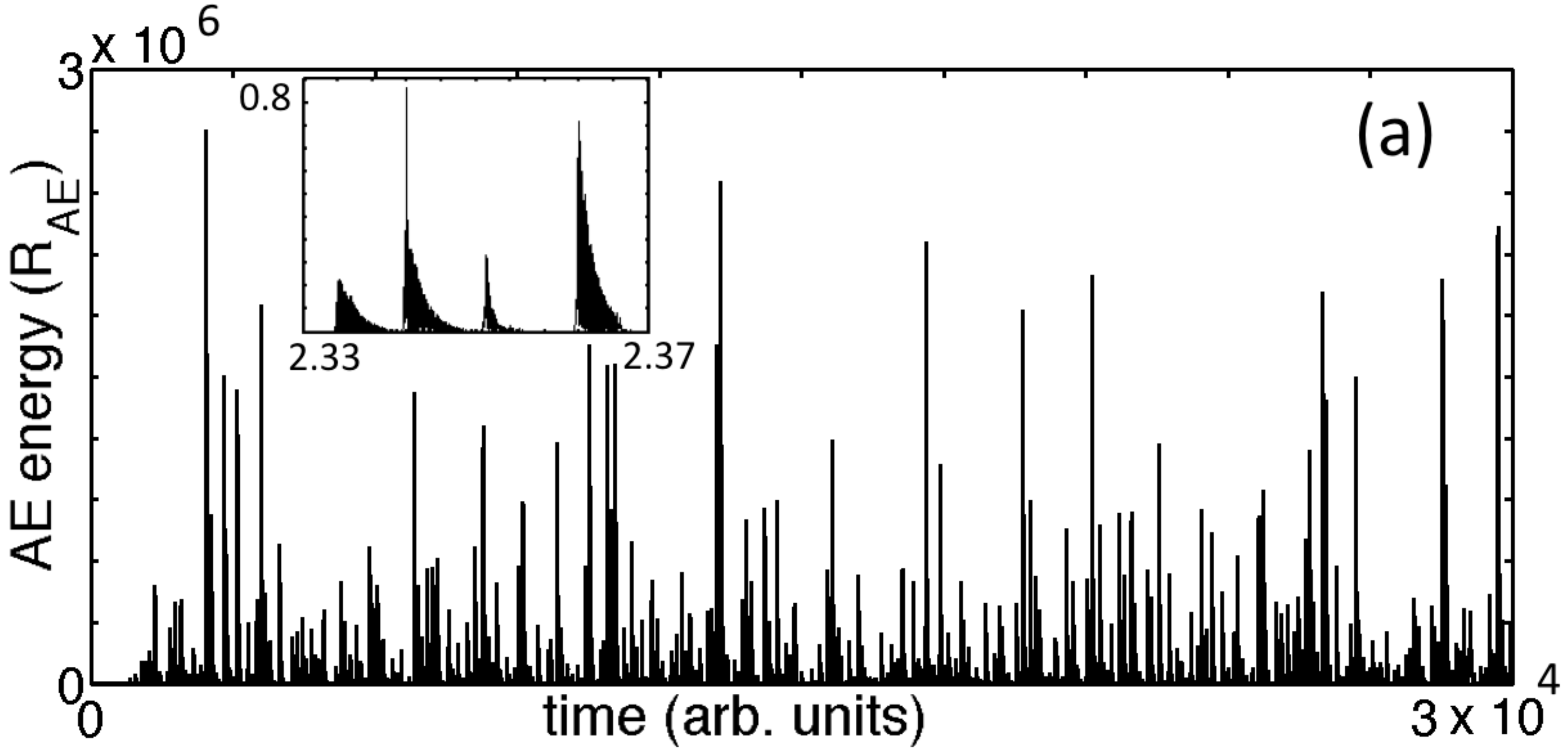}
\includegraphics[height=4.0cm,width=7.9cm]{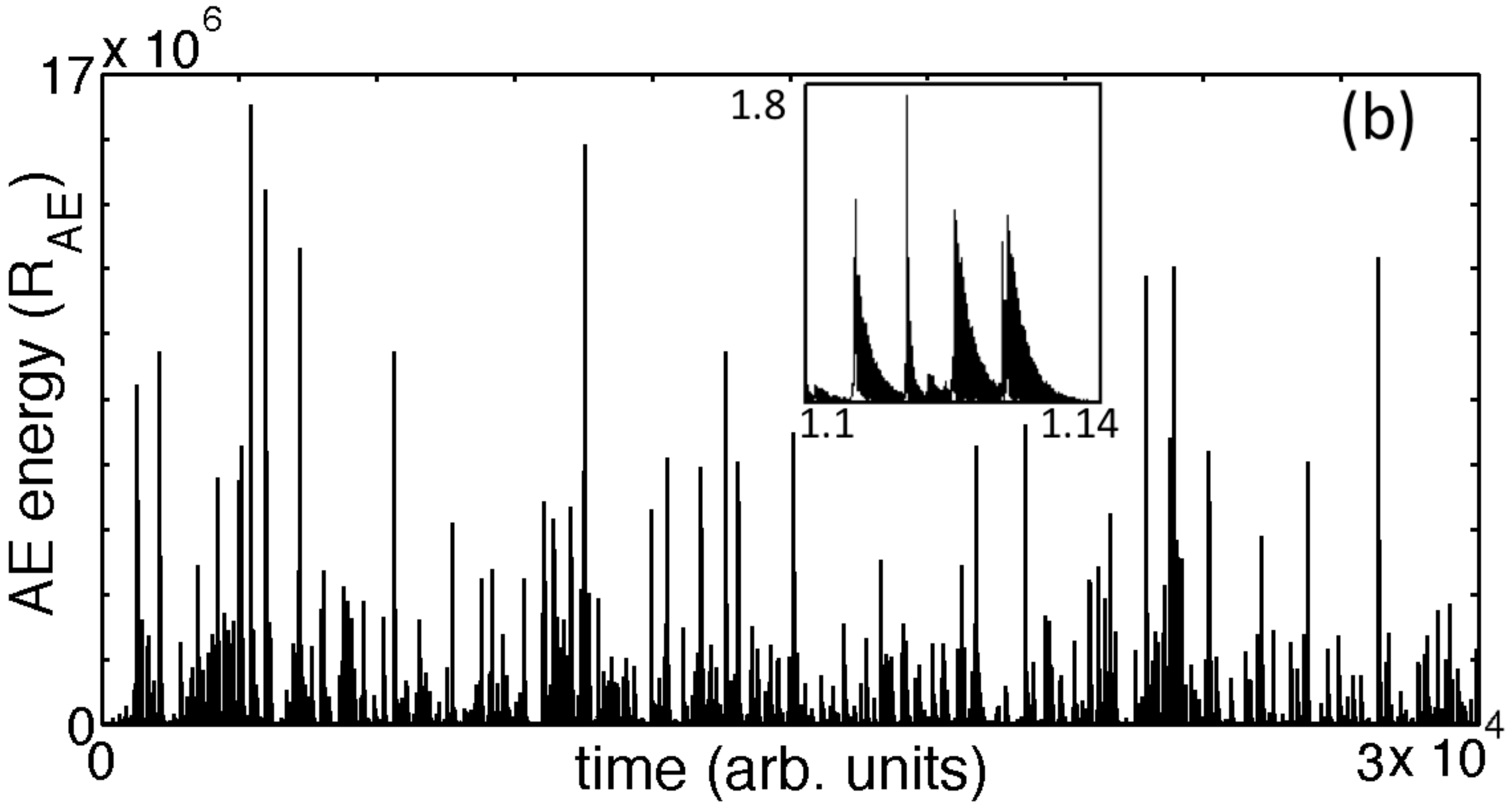}
\includegraphics[height=4.0cm,width=7.9cm]{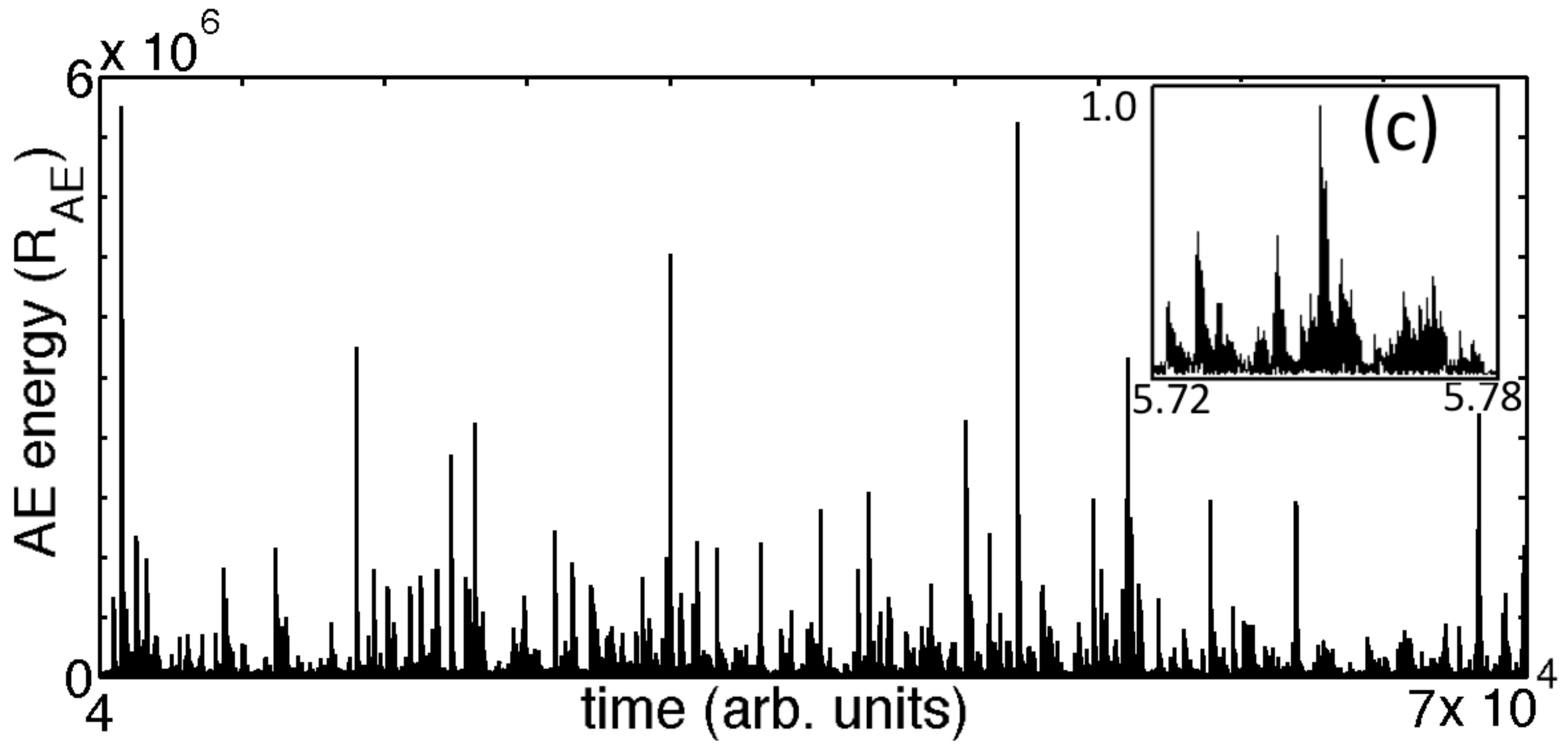}
\end{center}
}
\caption{Model acoustic energy   $R_{AE}$ associated with (a)  the uncorrelated static type C bands for $\dot\epsilon_a = 1.125 \times 10^{-5}/s$, (b) the partially propagating type B bands  for $\dot\epsilon_a = 3.0\times 10^{-5}/s$, and (c) the fully propagating type A bands for $\dot\epsilon_a = 5.5\times 10^{-5}/s$. The inset in (a) shows the  non-overlapping nature of  the bursts for the type C bands. The amplitude decreases in an oscillatory manner not visible on this scale.  The inset in (b) shows increasing background AE level  caused by SASs for the type B bands. The inset in (c) shows increased level of background AE is caused due to the long stretches of SASs for the type A bands.}
\label{RAE_BANDs}
\end{figure}

Now consider the nature of the AE signals  accompanying the three types of PLC bands. For the type C band, well separated AE bursts are seen in the strain rate region  $3 \times 10^{-6}/s < \dot \epsilon_a < 1.5 \times 10^{-5}/s$ \cite{Jag15}. Further, the bursts of AE are well correlated with the stress drops.  Figure \ref{RAE_BANDs}(a) shows a typical burst type AE signals  for $\dot\epsilon_a = 1.125\times 10^{-5}/s$. The inset shows a few successive well separated  AE bursts. The amplitude  of  each burst exhibits an oscillatory  exponential decay,  which however is not visible on the scale of  the inset, but  becomes visible on a finer scale.  These features are consistent with AE experiments for the type C bands  \cite{Chmelik02,Chmelik07}. 

With  increasing $\dot\epsilon_a$,  the AE bursts begin to overlap in the region of the partially propagating type B bands.   A typical plot of the AE  for $ {\dot \epsilon}_a =3\times 10^{-5}/s$  is  shown in Fig. \ref{RAE_BANDs}(b).  The overall structure of the AE pattern is similar to the voltage plot of experimental AE signals in Fig. 2 of  Ref. \cite{Shashkov-pl-12}. As shown in Ref. \cite{Jag15}, a careful analysis of the AE signals and stress serrations show  two features. First,  the low amplitude continuous AE signals are  well correlated with the band propagation induced SASs. Second,  the large amplitude AE bursts (shown in the inset) are well correlated with the large amplitude stress drops that  are identified with the nucleation of new bands or stoppage of bands.  This prediction is consistent with recent experimental studies on AE during the PLC bands \cite{Chmelik02,Chmelik07}. 

At high $\dot\epsilon_a$ of the fully propagating type A bands, the nature of the AE pattern becomes  nearly continuous, which again is due to the long stretch of SASs induced by type A bands propagating long distances.  A typical AE pattern for $\dot\epsilon_a = 5.5 \times 10^{-5}/s$ is shown in Fig. \ref{RAE_BANDs}(c). (It must be mentioned here that the small  strain rate range over which the instability is seen is the limitation of the AK model as used here which does not include the forest hardening term.  This term was primarily ignored for mathematical convenience.  However, inclusion of this term extends  the  instability range  to three orders in strain rate as in experiments \cite{RThesis}. ) As can be seen, the   AE pattern has nearly continuous background AE level with occasional relatively large  bursts. The continuous background AE is illustrated in the inset. The high background level of the model AE signals  corresponding to the type A band is also consistent with experiments \cite{Caceres87,Zeides90,Chmelik02,Chmelik07}.  In our earlier paper \cite{Jag15}, we have demonstrated  that the relatively large  AE bursts are correlated with the large amplitude stress drops caused due to either nucleation or  the band reaching the boundaries.  

\subsection{Power law distributions of  AE events associated with the PLC bands}
Accumulation of the  statistics of AE events require identifying a segment of  AE signals  as an 'individual AE  burst' or an AE event. Recall that the AE signals corresponding to the type C bands consists of well separated AE bursts of varying peak amplitudes. In addition, we also find overlap of several successive  AE bursts (see the inset of Fig. \ref{RAE_BANDs}(a)).   Each of these AE bursts shown in the inset of Fig. \ref{RAE_BANDs}(a) decay in an oscillatory fashion. While the  peak amplitude of all AE bursts decay in this manner,   a few bursts  may not relax fully due to the overlap with  the next AE burst. In such cases, the amplitude of the AE signal first decreases and then increases to a new peak amplitude. When the strain rate is increased, the tendency for overlap of AE bursts increases as is clear from the insets of Figs. \ref{RAE_BANDs}(b, c). This feature helps us to define  'an individual AE burst' or an 'AE event'  by the local peak amplitude of the burst. Using local peak amplitude of each burst as the event size, $R_{AE}(p)$,  we have computed the distribution of the event sizes $D(R_{AE}(p))$. If the distribution  $D(R_{AE}(p))$ follows a power law, then we have   
\begin{eqnarray}
D(R_{AE}(p)) \sim   R_{AE}(p)^{-\nu}
\label{event-size}
\end{eqnarray} 
where $\nu$ is the scaling  exponent. 
 
Here, it is useful to briefly summarize the method used for accumulating the  statistics of experimental AE bursts or events.  Several factors influence the statistics of the AE event sizes.   The first is   the threshold imposed to identify what is regarded as an 'individual AE burst'.  (See Fig. 6 of Ref. \cite{Lebymf-pl-13}.)  The threshold, in particular,  has a tendency to eliminate  the  AE signals that are smaller than the threshold value. In our calculation, the power law distributions  have been computed without any threshold. The second factor is that the exponent value depends on the region of the strain where the signal is recorded for the analysis. This  is related to fact that the stress-strain curves exhibit hardening.  In contrast,   the model  AE signals are recorded in the stationary region. 
 
\begin{figure}[t]
\vbox{
\begin{center}
\includegraphics[height=4.0cm,width=7.9cm]{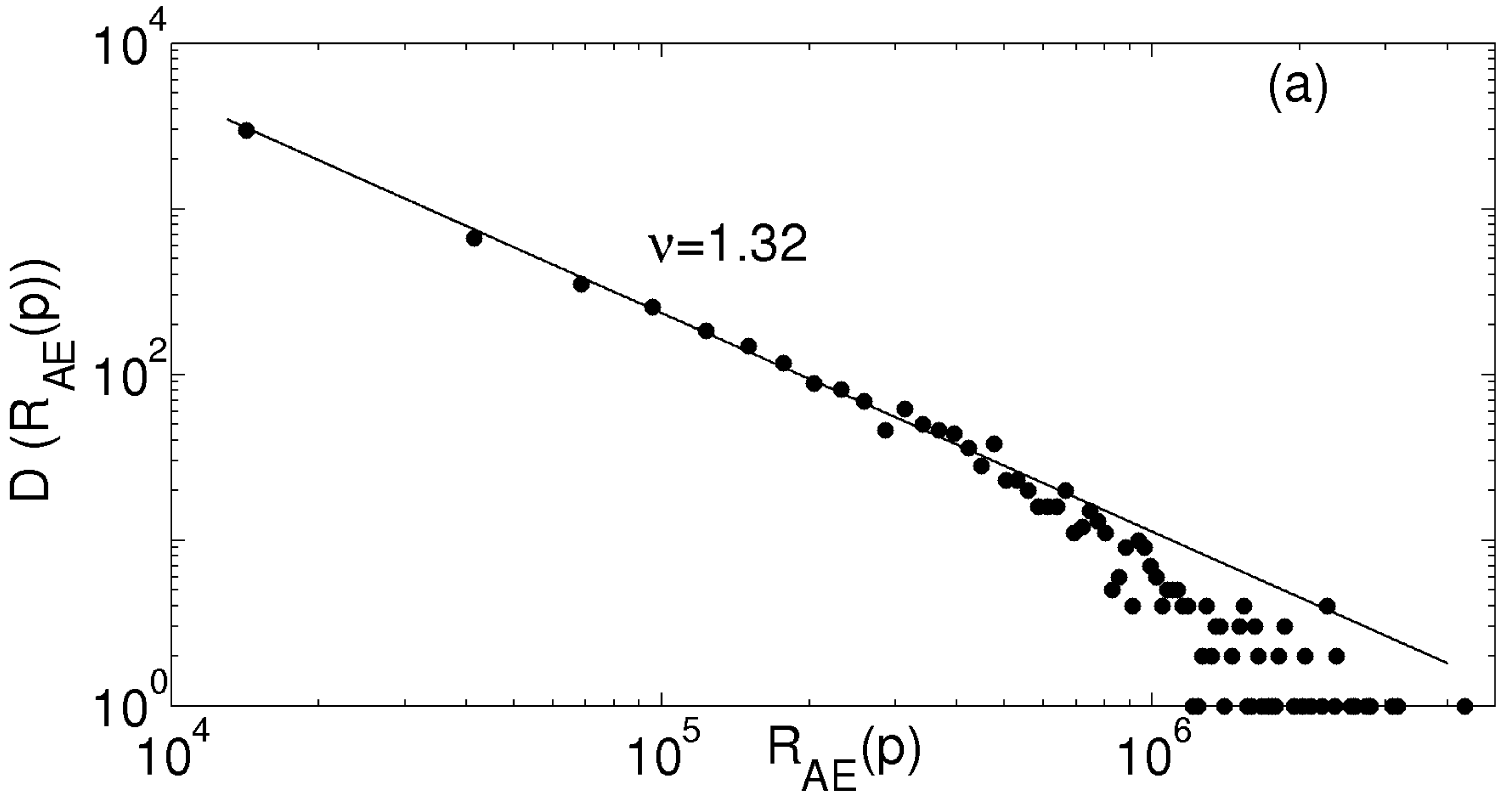}
\includegraphics[height=4.0cm,width=7.9cm]{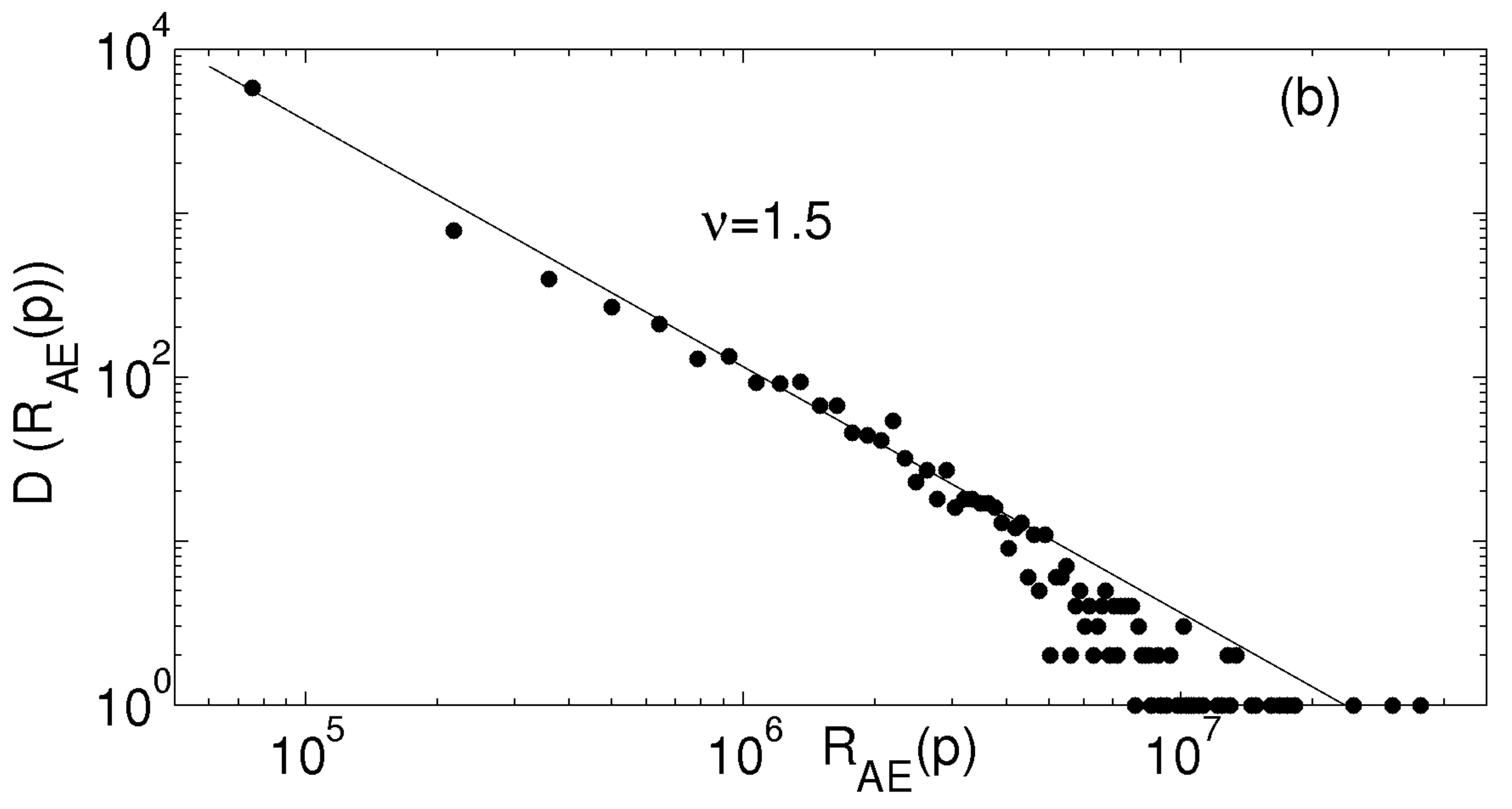}
\includegraphics[height=4.0cm,width=7.9cm]{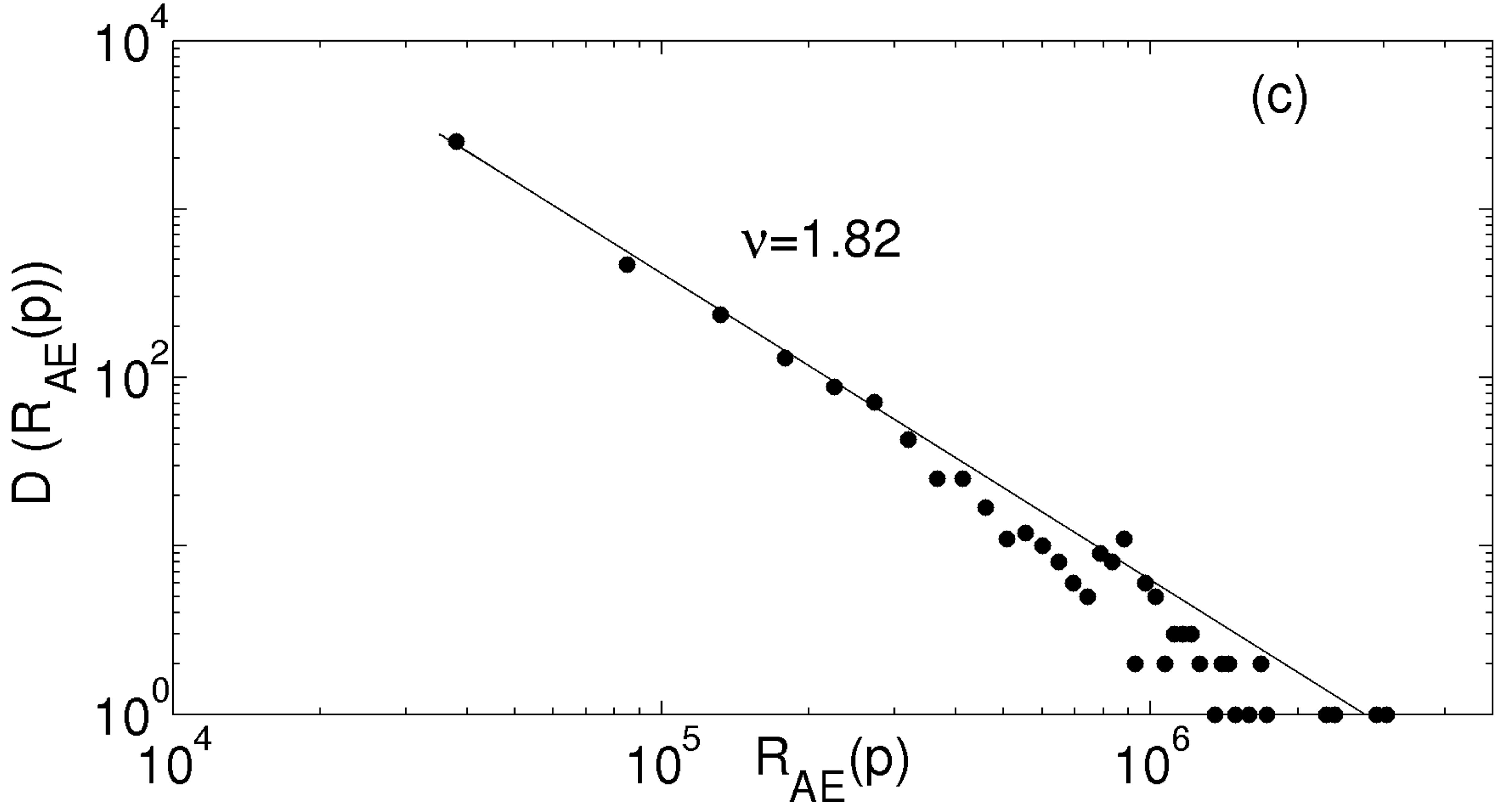}
\end{center}
}
\caption{Power law distributions for the AE events $R_{AE}(p)$ for (a)  the type C bands for  $\dot\epsilon_a = 1.125\times 10^{-5}/s$,   (b) the partially propagating type B band for  $\dot\epsilon_a = 3.0\times 10^{-5}/s$,  and (c) the type A band for $\dot\epsilon_a = 5.5\times 10^{-5}/s.$}
\label{PL-bands}
\end{figure}

We have calculated the distribution of the magnitude of the AE events (AE energy bursts)  $R_{AE}(p)$ associated with all  the three types of the PLC bands.   A plot of $log\, D(R_{AE}(p))$ versus $log\, R_{AE}(p)$  for the type C bands is shown in Fig. \ref{PL-bands}(a). A scaling region of nearly two orders of magnitude in $R_{AE}(p)$ is clear and the exponent value is $\nu = 1.32$.  At higher strain rates of partially propagating type B bands, the calculated  distribution $D(R_{AE}(p))$  is shown in Fig. \ref{PL-bands}(b). The scaling regime is again  nearly two orders in $R_{AE}(p)$ with an  exponent value $\nu =1.5$. In the fully propagating band A, we find significantly larger proportion of small amplitude AE events and smaller number of large amplitude AE events.   The corresponding power law distribution for $ R_{AE}(p)$ is shown in Fig. \ref{PL-bands}(c). As can be seen, there is a  reduced  scaling region,  a feature that is similar to experiments.  The exponent  value is $\nu =1.8$. In all the three cases, the model exponent values are considerably smaller than those reported by Lebyodkin  and co-workers \cite{Lebymf-Acta-12a,Shashkov-pl-12,Lebymf-pl-13}. 
\begin{figure}[!h]
\vbox{
\begin{center}
\includegraphics[height=4.0cm,width=7.9cm]{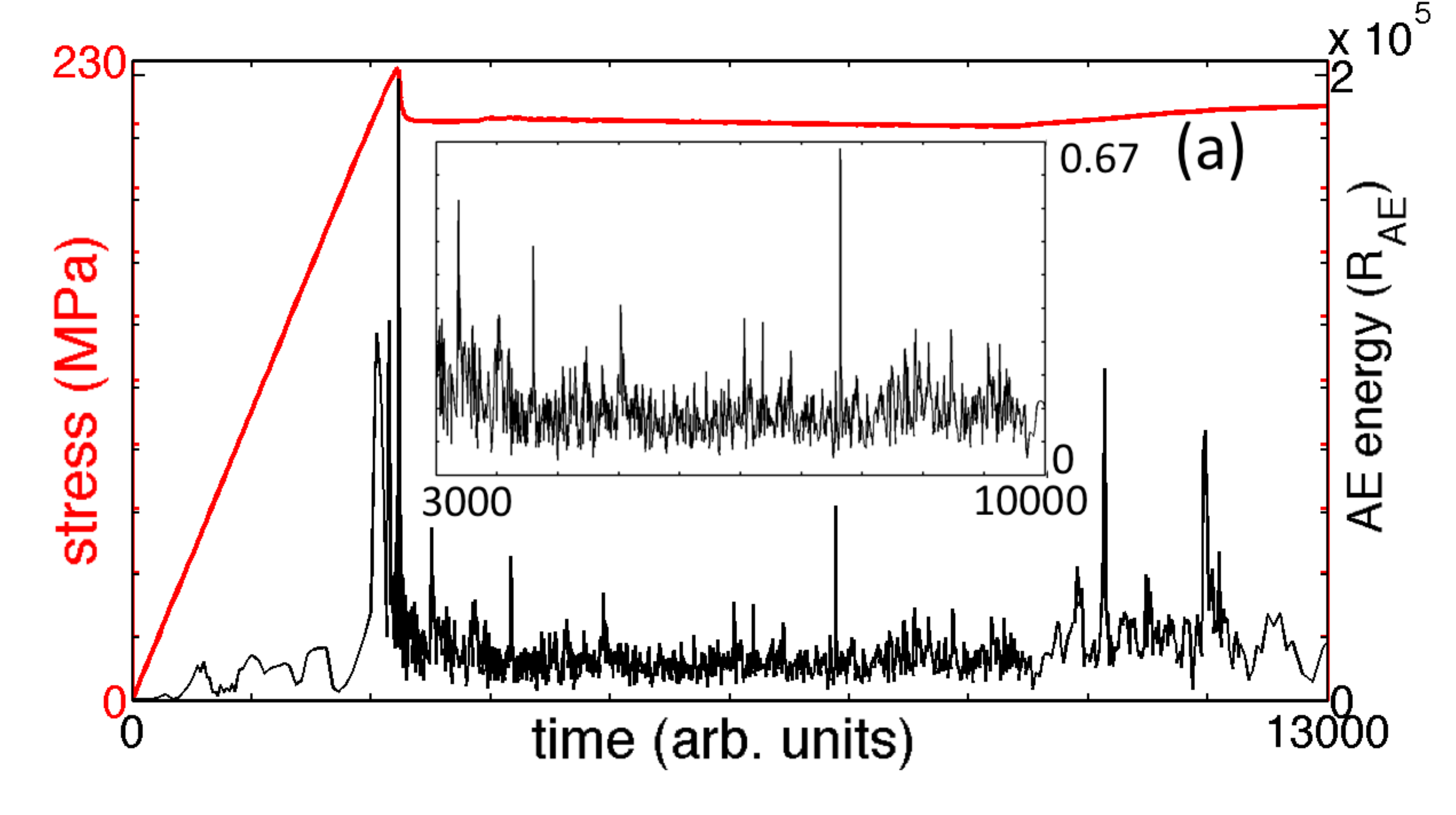}
\includegraphics[height=4.0cm,width=7.0cm]{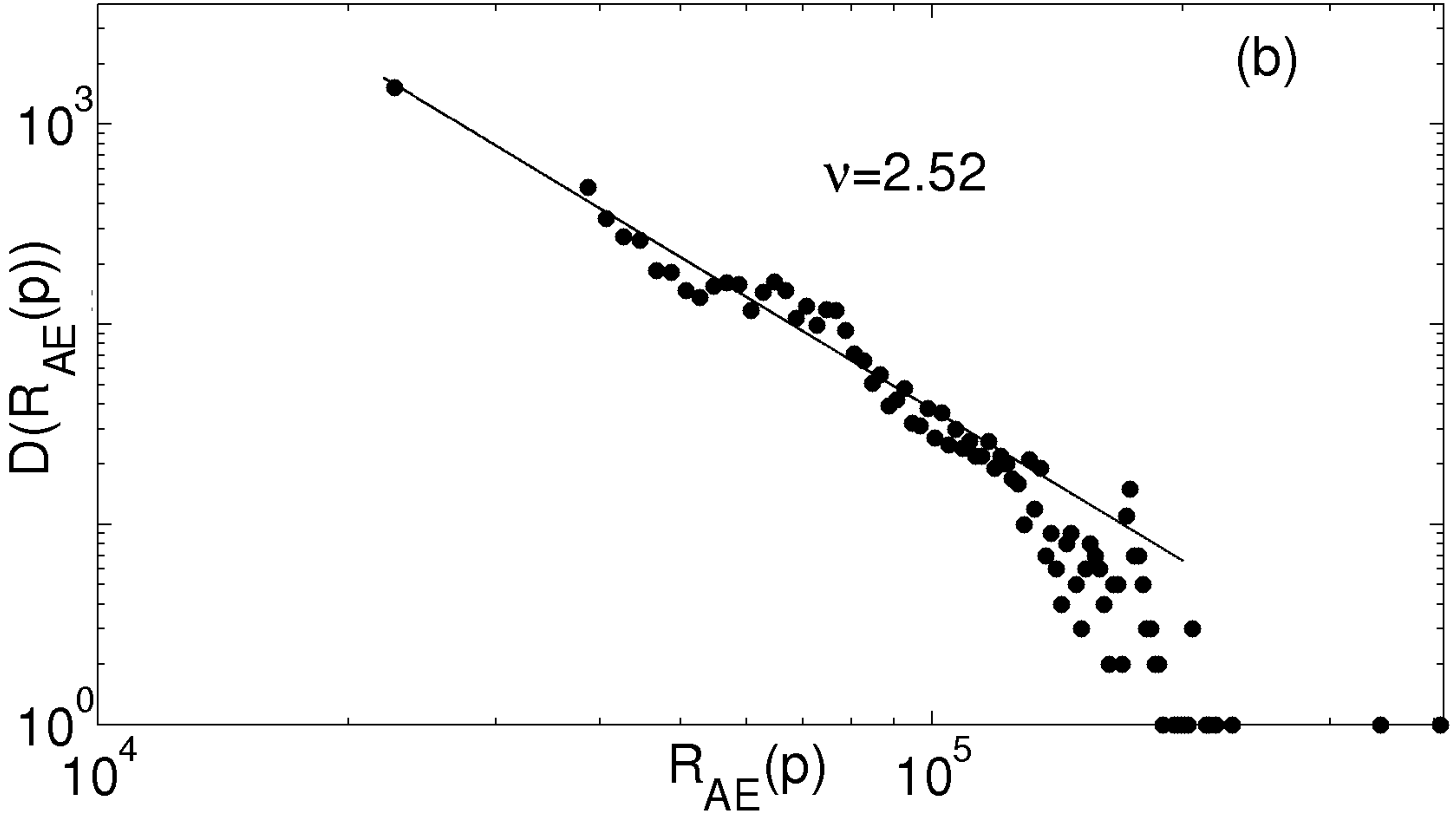}
\end{center}
}
\caption{(color online)  (a) Model acoustic energy $R_{AE}(p)$,  for the L{\"u}ders-like propagating band along with the stress-strain curve. (b) The corresponding power law distribution for the  AE signals, $\dot\epsilon_a = 1.67\times 10^{-6}/s.$}
\label{RAE_LUDERSBAND_B}
\end{figure} 

Finally, we consider  analyzing the AE signals  corresponding to another type of propagating band, namely, the L{\"u}ders band. It is well known that  several alloys exhibiting the PLC effect also exhibit L{\"u}ders band \cite{Chmelik02,Chmelik07,Zuev08}. Since the AK model predicts the characteristic features of the  three  PLC bands,  one can anticipate that the AK model should also  predict L{\"u}ders like band. Indeed,  the AK model  also predicts  L{\"u}ders like band following an yield drop. The corresponding  AE pattern has been investigated   \cite{Jag15}. Here,  we adopt the same  parameters as used in Ref. \cite{Jag15} for calculating the AE signals associated with  the  L{\"u}ders like band.  The values are: $\alpha_m=1/s,\alpha_c=0.002/s, \gamma= 5\times 10^{-4}/s, E^*/\sigma_y=240,m=10$ and $ \dot\epsilon_a=1.67 \times 10^{-6}/s$.  Figure \ref{RAE_LUDERSBAND_B}(a) shows the stress-strain curve and associated AE signals. As can be seen, the AE pattern  exhibits a peak corresponding to the yield drop beyond which the AE amplitude  decreases rapidly to an average AE level  corresponding  to the band propagation  regime.  The AE peak is due to the rapid multiplication of the dislocations from its initial low value. The steady level of  AE activity during the band propagation can be identified with the production of new dislocation sources at the propagating front as it moves along the sample. The low amplitude level of the AE signals is again attributable to the SASs induced by the propagating band (see the inset of Fig. 6(b) of Ref. \cite{Jag15}).  The nature of the AE pattern corresponding to the L{\"u}ders  band predicted by the AK model is  also consistent with experiments \cite{Chmelik02,Chmelik07,Zuev08}. The calculated distribution of the AE bursts (sampled in the band propagation region) is shown in Fig. \ref{RAE_LUDERSBAND_B}(b). As can be seen, the scaling region is only one order as for the  type A band. The  exponent $\nu$ $\sim 2.51$. This value is significantly higher than the model exponent values  for the type C, B  and A, but is closer to the type A propagating band. The latter is understandable since both are propagating type bands that have low level of stress fluctuations. However,  there are no reports of power law distribution for the experimental AE  signals in the case of L{\"u}ders band for comparison with the model exponent value. 

In summary, power law distributions are predicted for the three PLC bands with  exponents values increasing from $1.32$ for the type C band to $1.82$ for the type A band. Further, power law distribution  for  the L{\"u}ders band is also predicted by the model with an exponent  2.52.  Thus. the power law statistics of AE signals appears to be ubiquitous to PLC bands and the L{\"u}ders band.   

\section{Multifractal analysis}
\begin{figure}[t]
\vbox{
\begin{center}
\includegraphics[height=4.0cm,width=8.0cm]{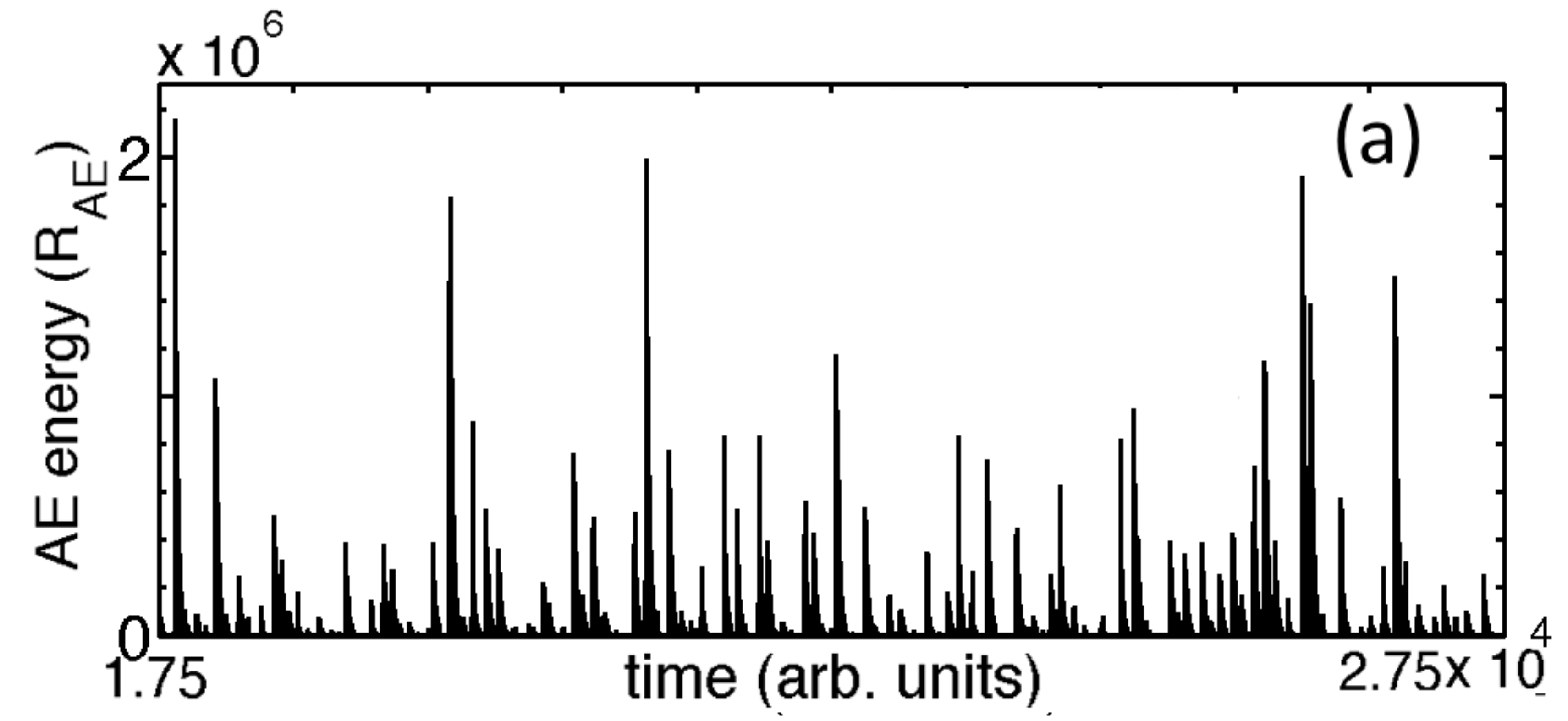}
\includegraphics[height=4.0cm,width=7.9cm]{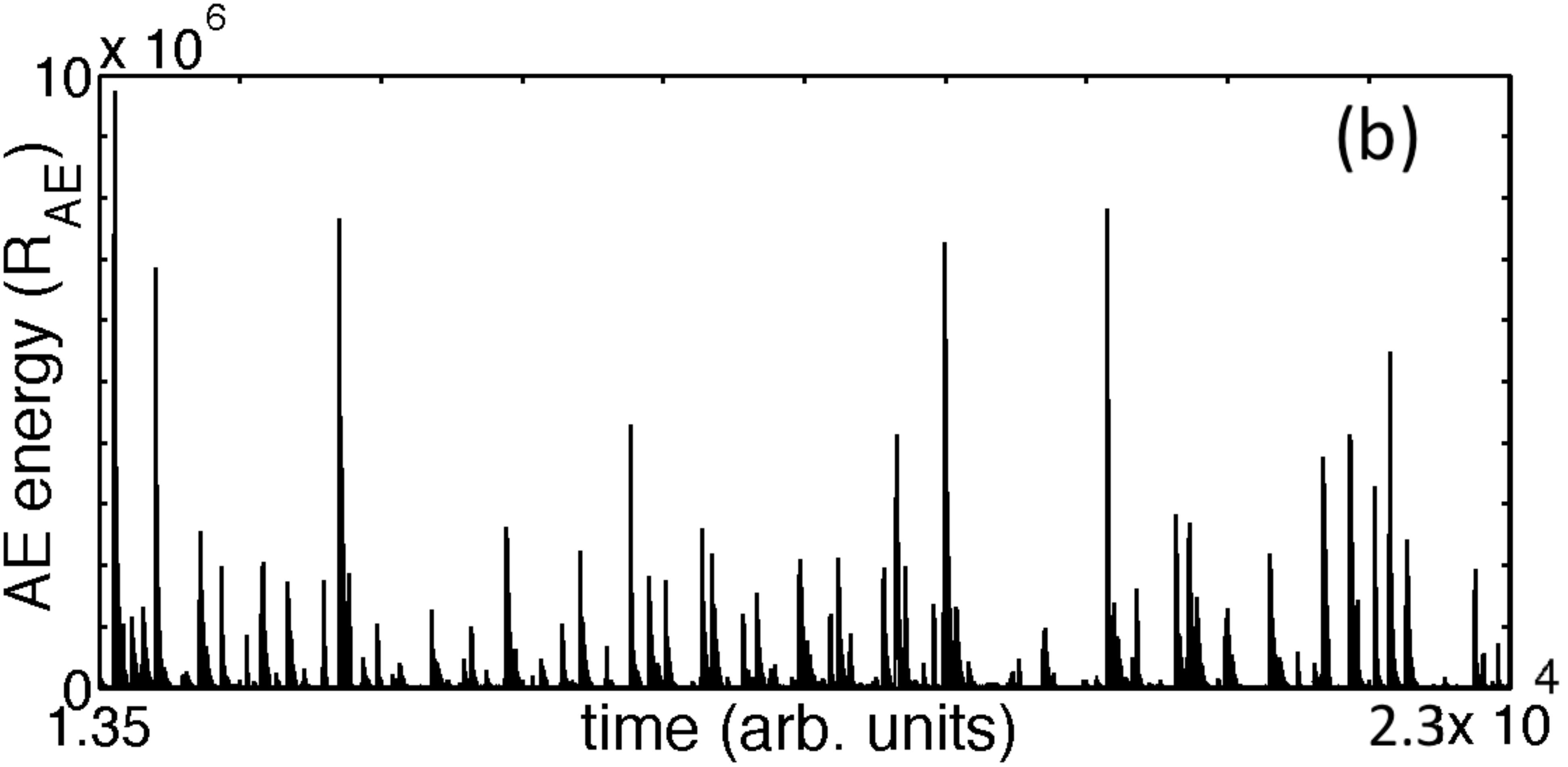}
\includegraphics[height=4.0cm,width=7.9cm]{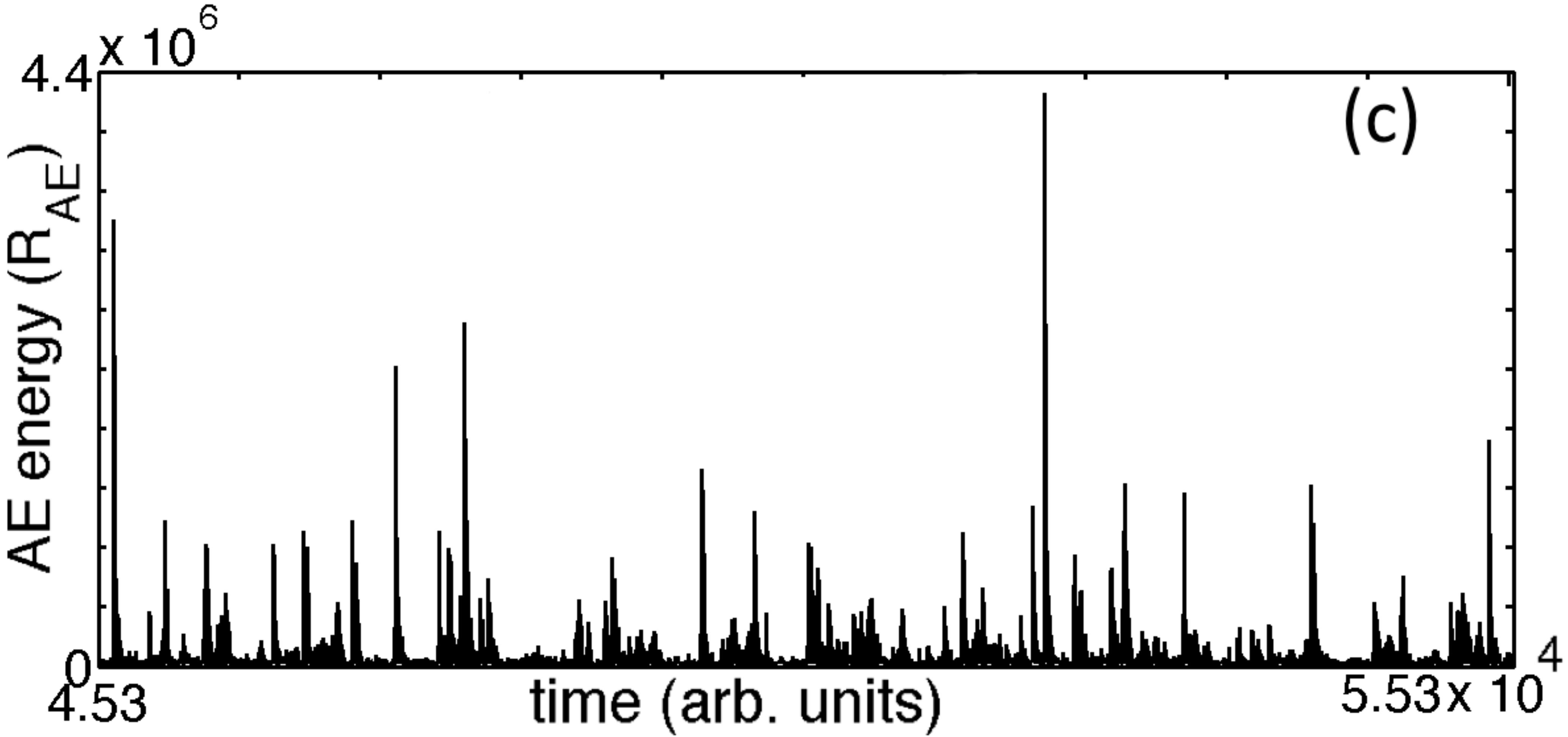}
\end{center}
}
\caption{Plots of respective  sub-segments of the AE signals  shown in (a)  Fig. \ref{RAE_BANDs}(a) for the  type C band. (b)    Fig. \ref{RAE_BANDs}(b)  corresponding to  the  type B band, and  (c)  Fig. \ref{RAE_BANDs}(c)  corresponding to the type A band. }
\label{Scale-sim}
\end{figure}

Scale invariant  power law distributions for events (of any kind) in a time series is the simplest statistical feature. Very often,  such time series can possess much richer scaling properties than power law distributions.   For instance,  the overall structure  of the AE pattern  corresponding to the three types of bands shown in Figs. \ref{RAE_BANDs} (a-c) is similar to  the dissipated energy pattern in turbulent flows (compare these figures with  Fig. 1 of Ref. \cite{Chhabra89}). In general, such time series can possess much richer scaling properties than power law distributions. This  can be visualized by examining different segments of the AE time series on smaller scales.  To illustrate this consider  the AE signals corresponding to $\dot\epsilon_a = 1.125\times 10^{-5}/s$ shown in Fig. \ref{RAE_BANDs}(a). The AE energy bursts do not appear to exhibit any obvious  correlation between the successive bursts. However, the AE time series is not random either because the magnitudes of the AE bursts obey a scale free power law distribution meaning that there is correlation at all scales. The time correlation is more subtle, which can be  visualized by plotting a sub-segment of the AE signals and comparing it with the whole. A plot of  the sub-interval   $t=1.75 - 2.75 \times 10^4$ is shown in Fig. \ref{Scale-sim}(a).  While a strict scale similarity with Fig. \ref{RAE_BANDs}(a) is not seen, the overall structure appears to be statistically similar to the original AE segment. In fact, this kind of statistical scale similarity is exhibited by any sub-segment. This suggests that  different segments have different scaling exponents.  This kind of statistical scale similarity is also seen for  the AE patterns corresponding to the type B and A bands. This is illustrated in the plots of sub-segments of Figs. \ref{RAE_BANDs}(b) and \ref{RAE_BANDs}(c)   shown in Fig. \ref{Scale-sim}(b) and \ref{Scale-sim}(c) respectively.   

Such multi-scale similarity of sets are known to lead to highly nonuniform probability distributions instead of a simple power law distribution. Such nonuniform distributions  require  a continuum of scaling exponents for their characterization.  The  emergence of such  nonuniform distributions has  been attributed to the underlying nonlinear dynamical evolution of the system as established in a number of physical situations such as turbulence, distribution of growth probabilities of a diffusion limited aggregate  \cite{Hents83,Fris84,Halsey86,Chhabra89,KRS91},  stress drop magnitudes  in type A PLC bands \cite{Anan99,Leby95,Bhar01} and  AE energies associated with  type C,  B and A bands    \cite{Lebymf-Acta-12a,Lebymf-Acta-12a}. Then, statistical  characterization of the self-similar (fractal) properties is carried out in terms of the  measures  associated with the nonuniform distribution.  This is the basis of multifractal formalism. The characterization  is done in terms of the commonly used generalized Renyi dimensions $D_q$   parametrized by $q$ or in terms of singularity spectrum $f(\alpha)$ of  singularity strengths $\alpha$  associated with the nonuniform distribution \cite{Renyi,Hents83,Fris84,Halsey86,Chhabra89}.  A direct way of calculating the  singularity spectrum was introduced as an alternate way of characterizing nonuniform distributions \cite{Chhabra89}.

\subsection{Generalized dimension and singularity spectrum}
As in many physical systems,  in our case also, the AE signals occur in time and therefore the support of the measure is the real line.   Then,  a length  $L$ of the time series can be covered by $N$ segments of  time interval  $\delta t$. Then, we  have $N \delta t = L$. Let, ${R_{AE}}_i$ be the acoustic energy dissipated in the $i^{th}$ interval. Then, the probability $P_i(\delta t)$ of the acoustic energy dissipated in the $i^{th}$ interval is given by $P_i(\delta t) = {R_{AE}}_i /\sum_{i}^N {R_{AE}}_i$. The  generalized dimension $D_q$ is defined by 
\begin{equation}
D_q = \frac{1}{q-1} \lim_{\delta t \rightarrow 0} \frac{ln \sum_iP_i^q}{ ln \, \delta t}.
\end{equation}
Here $q$ is a real number taken as a parameter. The structure of the above equation provides a straightforward physical interpretation. The positive $q's$ accentuate the denser regions (high probabilities) of the nonuniform distribution while the negative $q' s$ accentuate the rarer regions (small probabilities). 

Alternately, multifractals can be defined as  interwoven sets with  fractal (Hausdorff) dimensions $f(\alpha)$ having a singularity strength $\alpha$. In this formalism,  the  probability of the $i^{th}$ box is taken to scale as $P_i(\delta t) \sim \delta t^{-\alpha_i}$. Then, the number of boxes $N(\alpha)$  with singularity strength between $\alpha$ and $d\alpha$ is given by $N(\alpha)\sim \delta t^{-f(\alpha)}$.   The singularity spectrum  $f(\alpha)$ of a multifractal is then related to  $D_q$ through a Legendre transformation (provided $f(\alpha)$ and $D_q$ is a smooth function of $\alpha$ and $q$). Then, 
\begin{equation}
(q-1)D_q= q \alpha - f(\alpha) \, \, {\rm and } \, \, \alpha = \frac{d}{dq}(q-1) D_q.
\label{Dq-f}
\end{equation}
However,  obtaining the  $f(\alpha)$ spectrum using the computed $D_q$ values requires evaluation of the derivatives. This can  lead to uncontrolled errors, particularly in the analysis of  experimental  data. In view of this,  a direct computation of $f(\alpha)$ has been suggested \cite{Chhabra89}.  The method involves  defining normalized measures $ \mu_i(\delta t, q)$ by
\begin{equation}
\mu_i(\delta t, q) = \frac{P_i^q}{\sum_j  P_j^q}
\end{equation}
in terms of the probabilities $P_i$ defined earlier.  Using $\mu_i(\delta t,q)$ and $P_i(\delta t)$, we can directly calculate the multifractal spectrum $f(\alpha)$  as a function of $\alpha$ using \cite{Chhabra89} 
\begin{equation}
\alpha = \lim_{\delta t \rightarrow 0} {\frac {\sum_i \mu_i(\delta t, q) ln \,
P_i(\delta t)}{ln \, \delta t}},
\label{timalpha}
\end{equation}
and
\begin{equation}
f(\alpha) = \lim_{\delta t \rightarrow 0} {\frac {\sum_i \mu_i(\delta t, q)
ln \, \mu_i(\delta t, q)}{ln \, \delta t}}.
\label{timfalpha}
\end{equation}
\begin{figure}[h]
\begin{center}
\includegraphics[height=4.5cm,width=8.0cm]{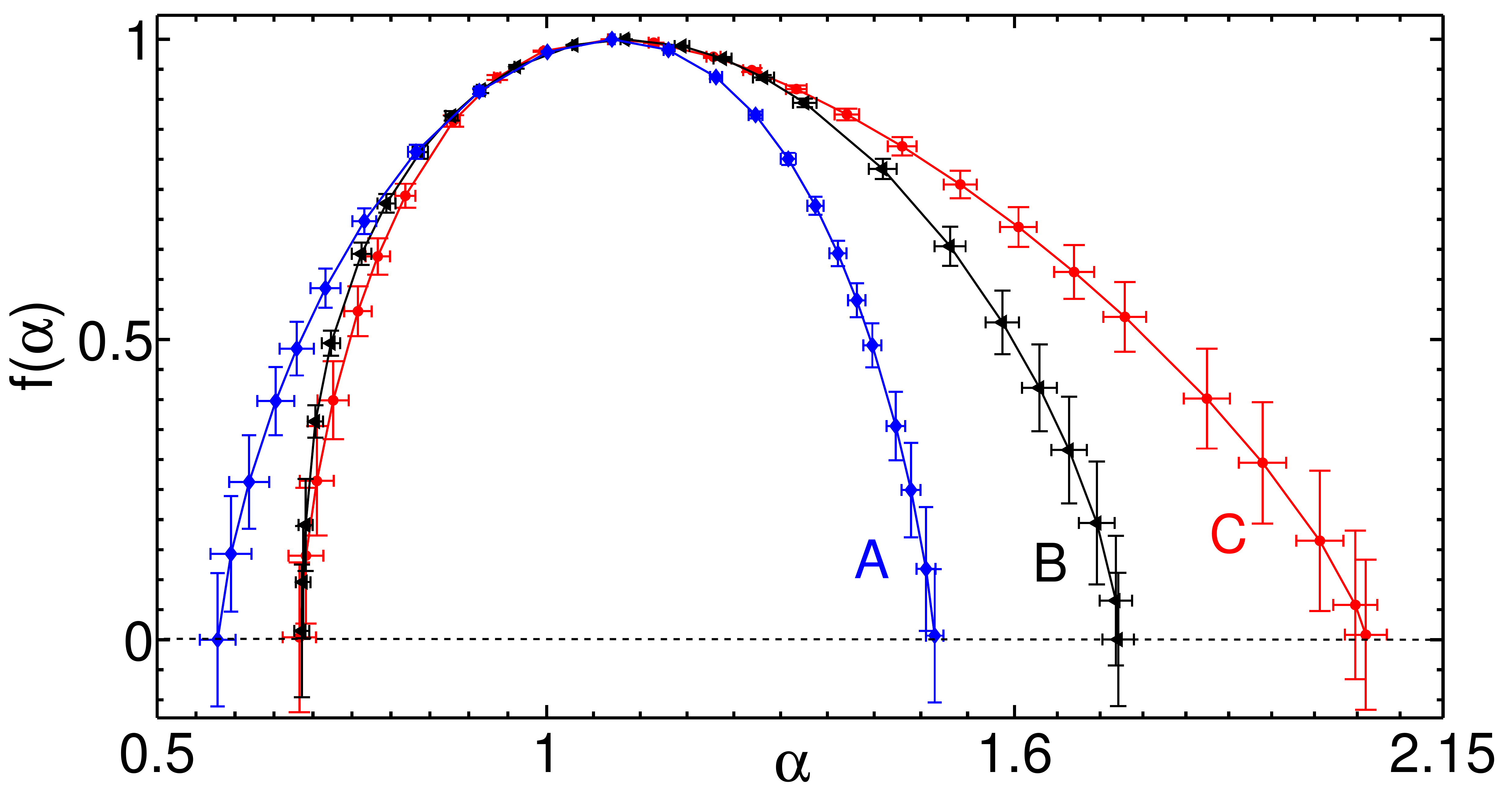}
\end{center}
\caption{(color online) Multifractal spectrum  $f(\alpha)$ associated with the AE signals   accompanying the type C, B and A bands for $\dot\epsilon_a= 1.125, 3.0 $ and $5.5 \times 10^{-5} s^{-1}$ respectively. }
\label{falpha-All-PLC}
\end{figure}
\begin{figure}[h]
\begin{center}
\includegraphics[height=4.5cm,width=8.0cm]{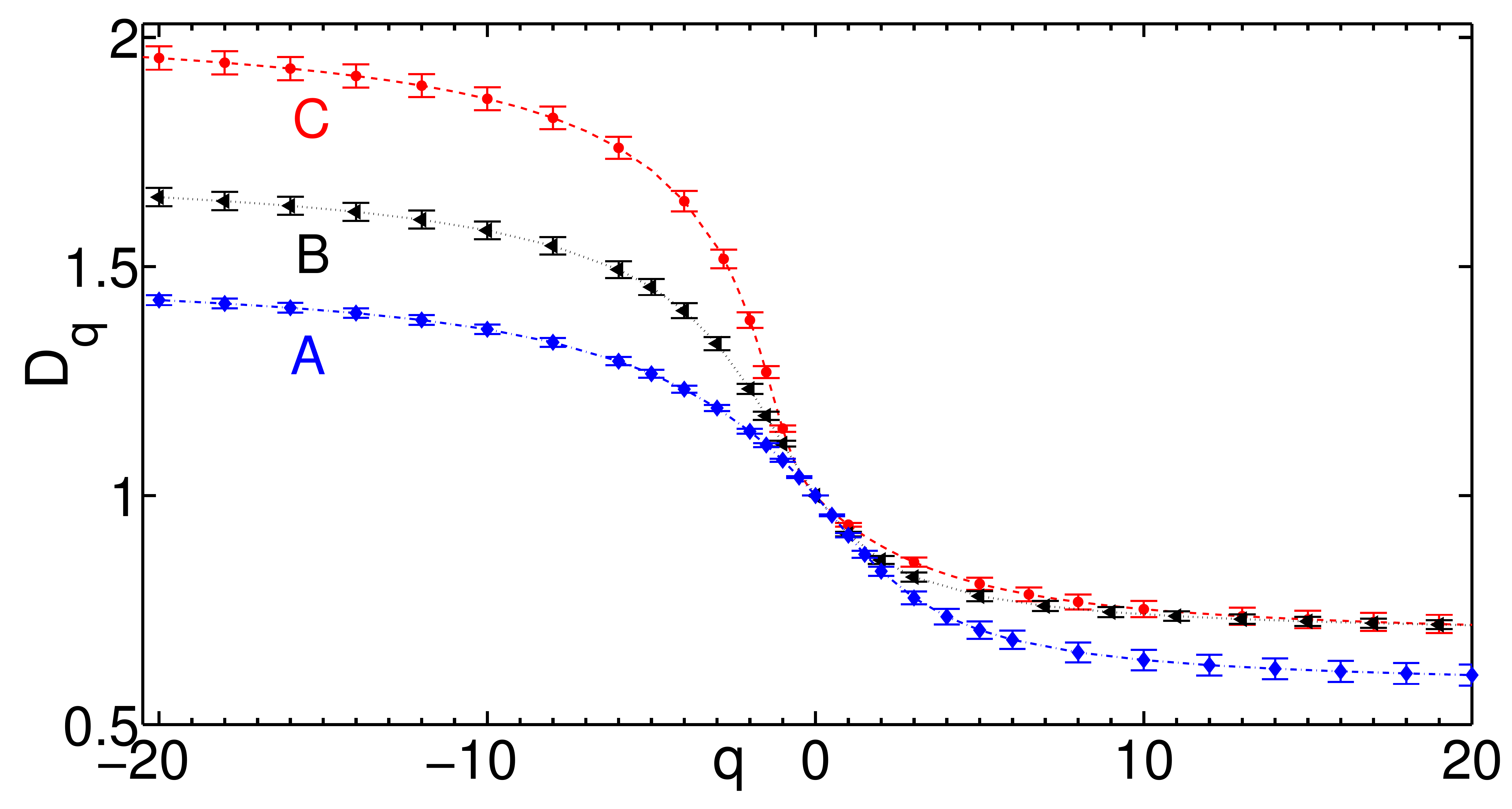}
\end{center}
\caption{(color online) Generalized dimensions for  the AE events associated with   the type C, B and A bands for $\dot\epsilon_a= 1.125, 3.0 $ and $5.5 \times 10^{-5} s^{-1}$. }
\label{Dq-All-PLC}
\end{figure}

We use the direct method of computing $f(\alpha)$ spectrum. Numerically, the multifractal spectrum for a given data set is calculated by plotting the  $log\, \sum \mu_i log \, \mu_i$  and $\log \, \sum \mu_i log P_i$ as a function of $log \,\delta t$ to obtain the respective slopes for each $q$. The slope of  $ log \, \sum \mu_i log \, \mu_i$ verses $log \, \delta t$ gives $f(q)$ while $log \, \sum \mu_i log \, P_i$ verses $log \, \delta t$ gives $\alpha(q)$. In general, the  log-log plots begin to exhibit scatter for increasing   positive and negative $q$ values. This is due to the poor statistics corresponding to small $\delta t$ bins. For this reason, a fit is obtained by considering those bins with reasonable statistics (i.e., bins of larger $\delta t$). Such a scatter  for large positive and negative values of $q$ are common to multifractal calculations. (See plots of  $\sum \mu_i log P_i$ in Fig. 6 of Ref. \cite{Chhabra89}.) In our calculation,  the range over which $\alpha(p)$ and $f(p)$ are calculated is {\it at least two orders of magnitude in $\delta t$. } While we had no difficulty in getting good fits to the slopes even for large $q$, the  error bars increase for large $q$,  particularly for $f(q)$,  as will be clear (see below).   

We have computed the  singularity spectrum $f(\alpha)$ corresponding to the AE spectra (shown in  Figs. \ref{RAE_BANDs}(a-c)) corresponding to type C, B and A bands.  Figure (\ref{falpha-All-PLC}) shows the $f(\alpha)$ spectra corresponding to the AE signals associated with the  three PLC bands. It is clear that the maximum value of $f(\alpha)$ is unity in all the three cases as it should be for a set whose support is the real line. Several features  are evident. First,  the $f(\alpha)$ spectrum of  the AE signals associated with the type C band is skewed to the right with significantly larger error bars for higher values of $\alpha$. This is due to higher proportion  of large amplitude AE bursts compared to the type B and A bands.   Second  the width of the multifractal spectrum defined by  $\theta$=$\alpha_{max}$-$\alpha_{min}$ with  $\alpha_{min}$ and $\alpha_{max}$ referring to  the  extreme values of $\alpha$, is maximum for the type C AE signals,  decreasing to  a minimum for the AE signals  corresponding to the type A band.  The third  feature is that   $\alpha_{min}$ is nearly the same for all the three band types. This feature is also understandable since  unlike the probability of large amplitude AE bursts  are quite different, the probabilities  for small amplitude AE signals are not significantly different. 

The generalized dimension corresponding the AE signals  for the three types bands can now be  easily calculated by using the Eq. (\ref{Dq-f}). The corresponding $D_q$'s  as a function of $q$ are shown in Fig. \ref{Dq-All-PLC}. Again, the range of $D_q$ is largest for the type C band while it is lowest for the type A band. 

\begin{figure}[h]
\begin{center}
\includegraphics[height=4.5cm,width=8.0cm]{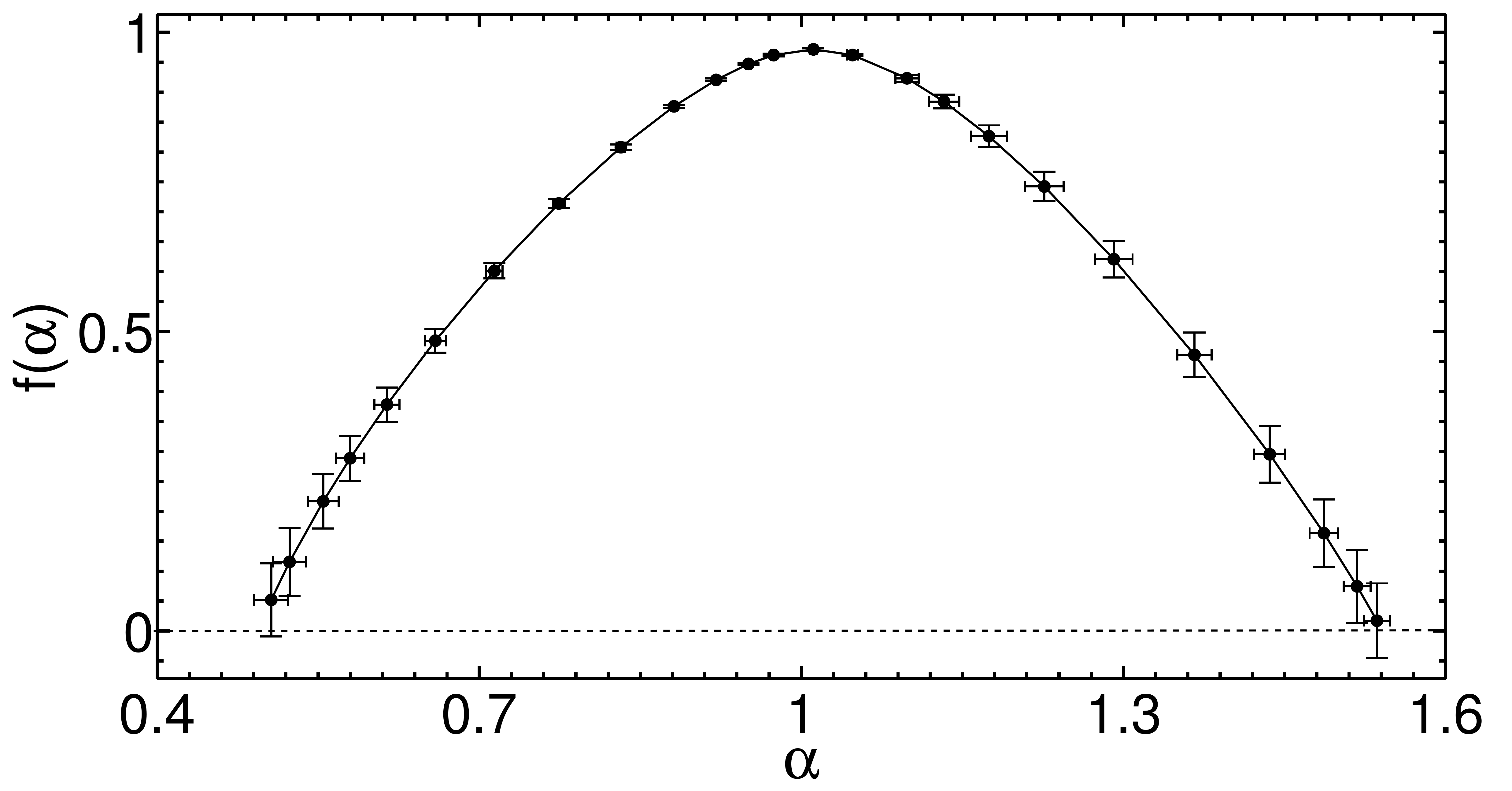}
\end{center}
\caption{Multifractal spectrum $f(\alpha)$ for the  L{\"u}ders-like propagating band for  $\dot\epsilon_a= 1.67 \times 10^{-6} s^{-1}$. } 
\label{falpha-Luders}
\end{figure}

We have also calculated multifractal spectrum for the L\"uders like band. This is shown in Fig. \ref{falpha-Luders}. The range of $\theta$=$\alpha_{max}$-$\alpha_{min}$ is similar to that of type A band as expected.
\section{Discussion and Conclusions}
The present work is motivated by lack of any model for the statistical characterization of the AE signals accompanying  the three band types in the PLC effect  \cite{Lebymf-pl-13,Leby-pl-Acta-12,Shashkov-pl-12,Lebymf-pl-13}. In particular, these studies report both power law distributions   and multifractal spectra for the AE  energies.  Two  types of statistical characterization of the model AE signals have been attempted. The first one is  to examine if the model AE signals follow   power law distributions for the  AE energies associated with  the C, B and A PLC bands. We have also carried out a similar study for  the L{\"u}ders  type band. The second type of statistical analysis is the possibility of multifractal spectra for  the AE energies associated with  the PLC and L{\"u}ders bands. Within our framework, the AE signals are calculated as solutions to the  discrete set of wave equations with plastic strain rate as a source term. The latter is computed using the AK model for the PLC effect since the  model predicts  the three PLC bands and the L{\"u}ders  like propagating band.  

The present study demonstrates that the statistical analysis of the  peak amplitudes of the model AE bursts associated with the three PLC bands  exhibit power law distributions  consistent with experiments \cite{Leby-pl-Acta-12,Shashkov-pl-12,Lebymf-pl-13}.  We have also verified that the  AE bursts   corresponding to the L{\"u}ders  like  band also follow a power law distribution.  Since there are no reports of the statistical analysis of the AE events for this case, it would be interesting to verify this prediction.  However, the model exponent values corresponding  the three PLC bands  are significantly smaller than the reported values.  Further, while the reported exponent values  decrease as we move from the type C to the type A bands, our results show the opposite trend. This mismatch  requires a critical examination.

In this context,  it must be stated that  a  comparison between the reported exponent values and the model exponent values is  not straightforward because the former depends on several experimental variables. Similarly, several factors such as modeling the AE signals and modeling the PLC bands  contribute to the computed exponent values. From the experimental side, there are several factors that affect the  exponent   such  as preparation of the samples, the nature of the sample, sample history, for instance, how long the samples are aged, sample-to-sample variation,  testing conditions etc.  From the modeling side, the one dimensional nature of the wave equation used for calculating the AE signals and  the calculation of the plastic strain rate source term clearly affect the exponent values. In addition, any effort  in modeling such a complex spatio-temporal phenomenon as the PLC effect can only constitute  idealization and therefore, can be expected to contribute.  Clearly, the AK model is no exception even though the model predicts all the characteristic features of the PLC effect mentioned in Section III A,  including the band types. 

While the above factors could be  contributing  to the smaller values  of  the model exponents  compared to the reported values,   the increasing trend of the model exponents  as we progress from type C to A bands needs a critical examination, particularly in view of the fact that the AK  model predicts most generic features of the PLC effect.  Here,  we argue that this trend  appears to be consistent with the physical (mathematical)   mechanism underlying acoustic emission used in our approach together with the dynamic strain aging mechanism specific to the PLC effect subsumed in the AK model. To see this, we first note that the model AE signals are computed by using the well established  mechanism that acoustic emission is triggered by  the release of the stored energy. The   latter information is contained in plastic strain rate and is  computed from the AK model. Now, we examine  how the aging kinetics of dislocations determines the nature of the AE pattern. For instance, the type C serrations are the result of unpinning of fully  aged  dislocations. This means that the unpinning stress is high. Thus, once unpinned, the kinetic energy imparted to the lattice is high and therefore  the  corresponding AE bursts are  generally  large. For higher strain rates where the type B and A  bands are seen, there is lesser time for aging process to complete and therefore  lower stress is required  to unpin dislocations compared to the type C bands. Consequently,  the transferred  kinetic energy to the lattice is lower leading to larger proportion of  lower level AE bursts. At high strain rates of the type A bands, there is even less time for aging process to occur, and therefore even smaller stress required to unpin dislocations. Therefore,  a large proportion of the AE signals should be expected to be  small amplitude signals and significantly fewer larger AE bursts. Indeed, we have recently demonstrated \cite{Jag15} that in general the AE bursts associated with type C are  large amplitude and well separated.  In contrast, for  the type B bands,  fluctuations in stress level are low during propagation with occasional large stress drops occurring during nucleation of a band or when the band stops \cite{Ritupan15}. In the type A band case,  due to the band propagating over large distances, the amplitude of the stress serrations are small with  relatively large stress drops seen only when the band reaches the sample boundaries.  Then,  using the fact that AE signals are proportion unpinning stress,  the relative  proportion of large  amplitude AE bursts are higher for the  type C than  for the type B or type A. This has been  demonstrated in Ref. \cite{Jag15}. On the other hand, considering the fact that relatively larger proportion of SASs are seen for the type B and even more for the A bands, we expect larger proportion of small amplitude AE bursts for the type B and A bands compared to the type C bands. This implies that the exponent corresponding type C band AE events should be lowest increasing for type B and then for the type A bands. This  is precisely the trend exhibited by the model exponents.

In view of the above discussion, we look for other possible sources of discrepancy between our model results and the experimental results.  For instance,  experimental stress-strain curves always exhibit considerable hardening. The  exponent value corresponding to the AE signals recorded in the regions of increasing strains increase from 2.5  approaching a  saturation level of 3.0. Noting that  signals can be considered as stationary only when the stress-strain curve reaches the saturation regime,  the AE time series itself is nonstationary during the  hardening region. This raises  questions about the possible influence of non-stationarity on the exponent values. This issue, however,  has  so far not been investigated in any context. Thus, the reported exponent values corresponding to the hardening regime are subject to this criticism. This is not applicable to the exponent value  corresponding to stationary regime ($\sim 3$).   Noting that the version of AK model used here exhibits low level of hardening,   even if  one wishes to examine the dependence on strain, the generalized AK model that  includes work hardening trend is more appropriate \cite{RThesis}.  In contrast,  the  model AE statistics has been compiled  in the stationary state. The model exponent values are however significantly low.

On the other hand, the exponent values are sensitive to the threshold value used for recording a AE burst as a single event. In experiments, a stretch of a AE signal is regarded as an individual AE event when  the amplitude of the signal stays above a chosen threshold value. Lebyodkin et al. \cite{Leby-pl-Acta-12} report that increasing the threshold by a factor of 3.3 decreases the exponent from 2.54 to 2.1 for the type C band  with a concomitant decrease in the scaling region (see Fig. 6 of Ref. \cite{Lebymf-pl-13}).  Further, higher threshold  has a tendency to eliminate small amplitude AE bursts thereby decreasing the exponent value. Indeed, we have verified that increasing the threshold value of recording,  the model  exponent value decreases along with the  scaling regime. Furthermore, for higher strain rates, smaller events dominate, two successive peak heights that are close to each other will be regarded as one when the threshold is higher. Thus, at high strain rates  small events are under-sampled. This suggests that the exponent values for the type B and A bands should be expected to be higher than what is reported. This means that using a lower  threshold can reverse the decreasing trend of the exponent values for the experimental AE signals as we increase strain rates. However, this depends on the capability of the experimental set-up. 

On the modeling side, other than the influence of idealizations discussed above, one obvious idealization that affects the calculated AE signals  is that the solutions of the model equations (both the wave equation and the AK model) have been  obtained in one  space dimension while real samples are three dimensional.  One other possible influence on the computed exponent values is the numerical accuracy of computation of the solutions of  the model differential equations.  We have specifically investigated this aspect by increasing the accuracy of computation by one order. We find that the exponent value increases, though not significantly. The second factor that could affect the model AE statistics is the algorithm employed for computing the AE signals.  Recall that AE  is calculated using  the plastic strain rate computed using the  AK model.   However, the computed plastic strain rate $\dot\epsilon_p$ has been obtained  using   Eq. (\ref{S-eqn}), which  assumes stress equilibration.  This was done for the sake of convenience of computation (and the  procedure akin to adiabatic methods). The  framework  itself is more general than the methodology followed here for computing $\dot\epsilon_p$. However, the stress equilibration constraint can easily be lifted by using the instantaneous stress that can be obtained from the elastic strain calculated from the wave equation.  This again can affect the exponent value.

In view of the above discussion,  it is unrealistic to expect a match of model exponent values with that reported for the experimental AE energies.  On the other hand, while factors that affect the exponents in experiments have a tendency to reverse  the increasing trend of the exponent values, the factors contributing from the modeling efforts have a tendency of reverse the increasing trend of model exponents. Again,  it is difficult to make a quantitative assessment of  how the experimental and theoretical factors   affect  the trends of the exponents, the  question  whether the exponent values should increase  with strain rate or not remains unresolved. 

We have also computed the multifractal spectra for  the AE signals associated with the three PLC bands and the L{\"u}ders like propagating band. The multifractal spectra have been  computed  using the direct method of computing the $f(\alpha)$ spectrum  \cite{Chhabra89}. This circumvents  using the Legendre transformation of $\tau(q)$ to obtain the $D(q)$, a procedure that  often leads to  uncontrolled errors. The direct method used here is better suited for analyzing numerical data.  The computed $f(\alpha)$ spectra for the three PLC bands are all smooth. The  width of the  multifractal spectrum decreases from a maximum value for the type C band to a minimum for the type A bands. The corresponding $D_q$'s also show the same trend. A comparison with the results reported by  Lebyodkin and co-workers \cite{Lebymf09,Lebymf-Acta-12a} is not possible since the reported $f(\alpha)$ spectra are not in the stationary state.   We have also calculated the  $f(\alpha)$ spectrum for the L{\"u}ders like band. There are no reports of the $f(\alpha)$ spectrum for the L{\"u}ders band for comparison. 

In summary, the significant characteristic feature, namely, the  power law distributions for  the AE bursts for all the three types of PLC band is predicted by our approach. However, the model exponent values and their increasing  trend  with increasing strain rates does not match with  reported values and trend.  The possible causes contributing these two differences discussed in detail show that a  comparison between model exponent values  and that for the experimental AE signals is  not meaningful because the reported exponent values depend on several experimental variables as well as those in modeling a complex spatio-temporal phenomenon such as the PLC effect.  Power law distribution for the model AE bursts is also obeyed for the  L{\"u}ders like band. The approach also correctly captures the reported multifractal nature of the AE spectra associated with the PLC bands and L{\"u}ders band.

\section*{ACKNOWLEDGMENTS}
GA wishes to acknowledge the financial support from the Board of Research in Nuclear Sciences Grant No. 2012/36/18-BRNS and the support from the Indian National Science Academy for the Honorary Scientist Position.  

\section*{Appendix}
\appendix
\renewcommand\thesection{A\Alph{section}}
\setcounter{equation}{0}
In this appendix, we briefly recall the discrete wave equations used for computing acoustic emission. See  Ref. \cite{Jag15} for details. To do this, consider a spring-block system of $N$ points of mass $m$ coupled to each other through near neighbor springs pulled at a constant strain rate by holding one end of the sample fixed and pulling  the other.   The fact that the ends of the sample are gripped by the machine translates to choosing the  force constant $k_m$ for the boundary points  to be  much larger than the spring constant $k_s$ for the interior points. Then, the kinetic energy $T$, the local potential energy $V_{loc}$, the gradient potential energy $V_{grad}$  and the dissipated acoustic energy $R_{AE}$ [Eq. (\ref{P-dissip})] are written down easily in terms of the displacement $u(i,t)$ defined at each site. It is then straightforward to set up the corresponding set of wave equations. Taking the spatial derivative of these equations, the discrete set of wave equations  in terms of the  elastic strain variables $\epsilon_e(i,t)$ that includes the plastic strain rate  source terms $\dot{\epsilon}_p(i,t)$   take the form 
\begin{eqnarray}
\label{waveqn_discret1}
& &\ddot \epsilon_e(1) = 0.0,\\
\nonumber
\label{waveqn_discret2} 
& &\ddot \epsilon_e(2) = -\frac{c^2}{a^2} \big[ \{\epsilon_e(2)-\epsilon_e(3)\}
 +\frac{k_m}{k_s}\epsilon_e(2) -\frac{\partial\dot{\epsilon}_p(2,t)}{\partial t} \\
\label{waveqn_discret3} 
& - & \frac{\eta'}{a^2\rho}\big[\dot \epsilon_e(2)-\dot \epsilon_e(3)\big]
+ \frac{D'}{a^4\rho}\big [\epsilon_e(4)+\epsilon_e(2)-2 \epsilon_e(3)\big], \\
\nonumber
& &\ddot \epsilon_e(3)= \frac{c^2}{a^2}\big[ \epsilon_e(4)+\epsilon_e(2)-2\epsilon_e(3)\big]-\frac{\partial\dot{\epsilon}_p(3,t)}{\partial t}\\
\nonumber
&+& \frac{\eta'}{a^2\rho} \{ \dot \epsilon_e(4)+\dot \epsilon_e(2)-2\dot \epsilon_e(3)\}\\
&-& \frac{D'}{a^4\rho} \{\epsilon_e(5)-4\epsilon_e(4)+ 5 \epsilon_e(3)- 2 \epsilon_e(2) \},\\
\nonumber
\label{waveqn_discret}
& &\ddot \epsilon_e(i)= \frac{c^2}{a^2} \{\epsilon_e(i+1)-2 \epsilon_e(i)+ \epsilon_e(i-1)\}- \frac{\partial\dot{\epsilon}_p(i,t)}{\partial t}\\
\nonumber
&+&  \frac{\eta'}{a^2\rho} \{ \dot \epsilon_e(i+1)-2\dot \epsilon_e(i) + \dot \epsilon_e(i-1)\}- \frac{D'} {a^4\rho}\big [\epsilon_e(i+2)\\
&-& 4 \epsilon_e(i+1)+ 6 \epsilon_e(i) - 4 \epsilon_e(i-1) + \epsilon_e(i-2)\big ],\\
\label{waveqn_discret99}
\nonumber
& &\ddot \epsilon_e(N-1)= - \frac{c^2}{a^2}\big[ \{\epsilon_e(N-1)-\epsilon_e(N-2)\}\\
\nonumber
&-& \frac{k_m}{k_s} \{\epsilon_e(N)-\epsilon_e(N-1)\}\big] -\frac{\partial\dot{\epsilon}_p(N-1,t)}{\partial t}\\
\nonumber
&+ &\frac{\eta'}{a^2\rho} \big[\dot \epsilon_e(N)+ \dot \epsilon_e(N-2)-2\dot \epsilon_e(N-1)\big] -\frac{D'}{a^4\rho} \big[\epsilon_e(N-3)\\
&-& 4\epsilon_e(N-2)+ 5 \epsilon_e(N-1)-2 \epsilon_e(N) \big].\\
\nonumber
\end{eqnarray}
The Eq. (\ref{waveqn_discret}) is valid for $i= 4$ to $N-1$. The mass density $\rho = m/a^3$ and the velocity of sound  is  $c= \sqrt{\mu/\rho}$ [see Eq. (\ref{comp_wave})] with $\mu=k_s/a, \eta'=\eta/a$ and $D'=Da$. The plastic strain rate $\dot\epsilon_p$ obtained from the AK model (\ref{X-eqn}-\ref{S-eqn}) is used as a source term in Eqs. (\ref{waveqn_discret1}-\ref{waveqn_discret99}). Equations  (\ref{waveqn_discret1}-\ref{waveqn_discret99}) are  solved by using differential equation solver (ode15s MATLAB solver) with appropriate initial and boundary conditions. The initial conditions are
\begin{equation}
\nonumber
\epsilon_e(1,0)=  0 ;   \epsilon_e(i,0) = 0 + \xi \times \epsilon_r, \quad i=2,..,N-1, 
\label{IC1} 
\end{equation}
where $\epsilon_r$ ($\sim 10^{-7}$) represents the strain due to inherent defects in the sample and $ \xi$ is  random number in the interval  $- \frac{1}{2} < \xi < \frac{1}{2}$.  The left hand side of the sample is fixed and right hand side is pulled at a constant strain rate $\dot\epsilon_a$.  So the boundary condition for the wave equation are
\begin{eqnarray}
\epsilon_e(1,t) = 0, \quad \epsilon_e(N,t) = \dot{\epsilon}_a t; 
\,\, t>0.
\label{BC1}
\end{eqnarray}
The time step required for integrating the equations (\ref{waveqn_discret1}-\ref{waveqn_discret99}) need to be substantially small compared to the AK model. This requires that the time variables in Eqs. (\ref{waveqn_discret1}-\ref{waveqn_discret99}) and Eqs. (\ref{X-eqn}-\ref{S-eqn}) are matched correctly. (Interpolated values for $\epsilon_p(k,t')$ are used as an input in Eqs. (\ref{waveqn_discret1}-\ref{waveqn_discret99})). ( For details see Ref. \cite{Jag15}.)


\begin{thebibliography}{99}

\bibitem{Dimiduk09} M. D. Uchic,  P. A. Shade and D. M. Dimiduk, Annu. Rev. Mater. Res. {\bf 39}, 361 (2009). %

\bibitem{Zaiser08} M. Zaiser, J. Schwerdtfeger, A. S. Schneider, C. P. Frick, B. G. Clark, P. A. Gruber and E. Arzt, Philos. Mag. {\bf88} 3861 (2008). %

\bibitem{Anan07} G. Ananthakrishna, Phys. Rep. {\bf 440}, 113 (2007).%

\bibitem{Yilmaz11} A. Yilmaz, Sci. Technol. Adv. Mater. {\bf 12}, 063001 (2011).%

\bibitem{Dunegan69} H. Dunegan and D. Harris, Ultrasonics {\bf 7}, 160 (1969). %

\bibitem{Han11} Z. Han, H. Luo and H. Wang,  Mater. Sci. Eng. A {\bf 528}, 4372 (2011).%

\bibitem{Weiss97} J. Weiss and J. R. Grasso, J. Phys. Chem. B,  {\bf 101}, 6113 (1997). %

\bibitem{Weiss07} J. Weiss, T. Richeton, F. Louchet, F. Chmel\'ik, P. Dobron, D. Entemeyer, M. Lebyodkin, T. Lebedkina, C. Fressengeas and R. J. McDonald, Phys. Rev. B {\bf 76}, 224110 (2007). %

\bibitem{Miguel01} M. C. Miguel, A. Vespignani, S. Zapperi, J. Weiss and J. R. Grasso, Nature (London) {\bf 410}, 667 (2001). %

\bibitem{Lebymf-pl-13} M. A. Lebyodkin, I. V. Shashkov, T. A. Lebedkina, K. Mathis, P. Dobron, and F. Chmel\'ik, Phys. Rev. E, {\bf 88}, 042402 (2013). %

\bibitem{James71} D. R. James and S. H. Carpenter, J. App. Phys. {\bf 42}, 4685 (1971). %

\bibitem{Caceres87} C. H. Caceres and Rodriguez, Acta Metall. {\bf 35}, 2851 (1987). %

\bibitem{Zeides90} F. Zeides and J. Roman, Scr. Metall. Mater {\bf 24}, 1919 (1990). %

\bibitem{Chmelik02} F. Chmel\'ik, A. Ziegenbein, H. Neuha\"user and P. Luk\'a\v c, Mater. Sci. Eng. A {\bf 324}, 200 (2002). %
 
\bibitem{Chmelik07} F. Chmel\'ik, F. B. Klose, H. Dierke, J. \v Sachl, H. Neuh\"auser and P. Luk\'a\v c, Mater. Sci. Eng. A {\bf 462}, 53 (2007). %

\bibitem{Zuev08} T. V. Murav'ev and L. B. Zuev, Technol. Phys. {\bf 53}, 1094 (2008).%

\bibitem{Shibkov11} A. A. Shibkov and A. E. Zolotov, Crystalogra. Rep. {\bf 56}, 141 (2011).%

\bibitem{Lebymf09} M. A. Lebyodkin, T. A. Lebedkina, F. Chmel\'ik, T. T. Lamark, Y. Estrin, C. Fressengeas and J. Weiss, Phys. Rev. B {\bf 79}, 174114 (2009). %

\bibitem{Lebymf-Acta-12a}  M. A. Lebyodkin, N. P. Kobelev, Y. Bougherira, D. Entemeyer, C. Fressengeas, T. A. Lebedkina, I. V. Shashkov, Acta Mater. {\bf 60}, 844 (2012). %

\bibitem{Jag11} Jagadish Kumar and G. Ananthakrishna, Phys. Rev. Lett. {\bf 106}, 106001 (2011). %

\bibitem{Jag15} Jagadish Kumar, Ritupan Sarmah, and G. Ananthakrishna, Phys. Rev. B {\bf 92}, 144109 (2015). %

\bibitem{Neuhausser83} H. Neuh$\ddot a$user, in {\it Dislocations in Solids}, Vol. 6, edited by F. R. N. Nabarro, (North Holland, Amsterdam, 1983). %

\bibitem{Leby-pl-Acta-12}  M. A. Lebyodkin, N. P. Kobelev, Y. Bougherira, D. Entemeyer, C. Fressengeas, V. S. Gornakov, T. A. Lebedkina, I. V. Shashkov, Acta Mater. {\bf 60}, 3729 (2012). %

\bibitem{Shashkov-pl-12} I. V. Shashkov, M. A. Lebyodkin and T. A. Lebedkina, Acta Mater. {\bf 60}, 6842 (2012).%

\bibitem{Diodati91} P. Diodati, F. Marchesoni and S. Piazza, Phys. Rev. Lett. {\bf 67}, 2239 (1991). %

\bibitem{Petri94} A. Petri, G. Paparo, A. Vespignani, A. Alippi, and M. Costantini, Phys. Rev. Lett. {\bf 73}, 3423 (1994). %
 
\bibitem{Planes95} E. Vives, I. R\`afols, L. Ma\~nosa, J. Ort\'in, and A. Planes, Phys. Rev. B {\bf 52}, 12644 (1995).%

\bibitem{Rajeevprl} R. Ahluwalia and G. Ananthakrishna, Phys. Rev. Lett. {\bf 86}, 4076 (2001). %

\bibitem{Kalaprl} S. Sreekala and G. Ananthakrishna, Phys. Rev. Lett. {\bf 90}, 135501 (2003).%

\bibitem{Cicc04} M. Ciccotti, B. Giorgini, D. Villet, and M. Barquins, Int. J. Adhes. Adhes. {\bf 24}, 143 (2004). %

\bibitem{Rumiprl} Rumi De and G. Ananthakrishna, Phys. Rev. Lett. {\bf 97}, 165503 (2006). %

\bibitem{Jag08} Jagadish Kumar, Rumi De and G. Ananthakrishna, Phys. Rev. E {\bf 78}, 066119 (2008). %

\bibitem{Anan99} G. Ananthakrishna, S. J. Noronha, C. Fressengeas, and L. P. Kubin, Phys. Rev. E {\bf 60}, 5455 (1999). %

\bibitem{Leby95}  M. A. Lebyodkin, Y. Brechet, Y. Estrin and L. P. Kubin, Phys. Rev. Lett. {\bf 74},  4758 (1995). %

\bibitem{Bhar-Acta-02} M. S. Bharathi, M. A. Lebyodkin, C. Fressengeas and L. P. Kubin, Acta Mater. {\bf 50}, 2813 (2002). %

\bibitem{Csikor07} F. F. Csikor C. Motz, D. Weygand, M. Zaiser, S. Zapperi, Sicence, {\bf 318}, 251 (2007). %

\bibitem{Bak88} P. Bak, C. Tang, and K. Wiesenfeld, Phys. Rev. Lett. {\bf 59}, 381 (1987); Phys. Rev. A {\bf 38}, 364 (1988). %

\bibitem{Renyi} A. Renyi, {\it Probability Theory} (North-Holland, Amsterdam, 1970). %

\bibitem{Halsey86} T. C. Halsey, M. H. Jensen, L. P. Kadanoff, I. Procaccia, and Boris I. Shraiman, Phys. Rev. A {\bf 33}, 1141 (1986).%

\bibitem{Chhabra89} A. B. Chhabra and R. V. Jensen, Phys. Rev. Lett. {\bf 62}, 1327 (1989); A. B. Chhabra, C. Meneveau, R. V. Jensen and K. R. Sreenivasan, Phys. Rev. A {\bf 40}, 5284 (1989). %

\bibitem{Anan82} G. Ananthakrishna and M.C. Valsakumar, J. Phys. D {\bf 15}, L171 (1982). %

\bibitem{Bhar02} M. S. Bharathi and G. Ananthakrishna, Europhys. Lett. {\bf 60}, 234 (2002).%

\bibitem{Bhar03} M. S. Bharathi and G. Ananthakrishna, Phys. Rev. E {\bf 67}, 065104(R) (2003). %

\bibitem{Bhar03a} M. S. Bharathi, S. Rajesh and G. Ananthakrishna, Scr. Mater. {\bf 48}, 1355 (2003). %

\bibitem{Anan04} G. Ananthakrishna and M. S. Bharathi, Phys. Rev. E {\bf 70}, 026111 (2004). %

\bibitem{Ritupan15} Ritupan Sarmah and G. Ananthakrishna, Acta Mater. {\bf 91}, 192 (2015).  %

\bibitem{Rumiepl} Rumi De and G. Ananthakrishna, Europhys. Lett. {\bf 66}, 715 (2004). %

\bibitem{Land} L. D. Landau and E. M. Lifschitz, {\it Theory of Elasticity} (Pergamon, Oxford, 1986). %

\bibitem{Ritupan14} R. Sarmah and G. Ananthakrishna, Commun. Nonlinear Sci. Numer. Simulat., {\bf 19}, 3880 (2014).

\bibitem{Rajesh} S. Rajesh and G. Ananthakrishna, Phys. Rev. E {\bf 61}, 3664 (2000). %

\bibitem{Anan83} G. Ananthakrishna and M. C. Valsakumar, Phys. Lett. A {\bf 95}, 69 (1983).

\bibitem{Noro97} S. J. Noronha, G. Ananthakrishna, L. Quaouire, C. Fressengeas and L. P. Kubin, Int. J. Bifurcation Chaos Appl. Sci. Eng. {\bf 7}, 2577 (1997). %

\bibitem{Bhar01} M. S. Bharathi, M. Lebyodkin, G. Ananthakrishna, C. Fressengeas, and L. P. Kubin, Phys. Rev. Lett. {\bf 87}, 165508 (2001). %

\bibitem{Ranc05} N. Ranc and D. Wagner,  Mater. Sci. Eng. A {\bf 394}, 87 (2005). %

\bibitem{Jiang07} H. Jiang , Q. Zhang, X. Chen,  Z. Chen, Z.  Jiang  and X. Wu,  Acta Mater. {\bf 55}, 2219 (2007). %

\bibitem{RThesis} Ritupan Sarmah, Ph. D thesis, Indian Institute of Science,
Bangalore, India, 2012.

\bibitem{Hents83} H. G. E. Hentschel and I. Proccacia, Physica {\bf 8D}, 435 (1983).%

\bibitem{Fris84} U. Frisch and G. Parisi in {\it Turbulence and Predictability of Geophysical Flows and Climate Dynamics}, Ed.  M. Ghil, R. Benzi, and G. Parisi (North-Holland, New York, 1985), p.84. %

\bibitem{KRS91} K. R. Sreenivasan, Annu. Rev. Fluid. Mech. {\bf 23}, 539 (1991).

\end{thebibliography}
\end{document}